\documentclass[a4paper,11pt]{article}
\usepackage{amsmath}
\usepackage{color}
\usepackage{graphicx}
\usepackage{amssymb}
\usepackage{float}
\usepackage{scrtime}
\usepackage{fancyhdr}
\usepackage{subfigure}
\usepackage{authblk}
\usepackage{lipsum}
\usepackage{rotating}
\usepackage{floatpag}
\usepackage{varioref}
\usepackage{slashed}
\usepackage{enumerate}
\usepackage[toc,page]{appendix}

\usepackage[colorlinks=true
,urlcolor=blue
,anchorcolor=blue
,citecolor=blue
,filecolor=blue
,linkcolor=black
,menucolor=blue
,pagecolor=blue
,linktocpage=true
]{hyperref}

\usepackage[numbers,sort&compress]{natbib}

\setlipsumdefault{131}

\newcommand{\maton}{\texttt{maton}}

\newcommand{\susyflavor}{\texttt{susy\_flavor}}

\newcommand{\micromegas}{\texttt{MicrOmegas}}

\newcommand{\tbtau}{\ensuremath{t-b-\tau}}
\newcommand{\btau}{\ensuremath{b-\tau}}
\newcommand{\tanb}{\ensuremath{\text{tan} \beta}}

\newcommand{\SO}[1]{\ensuremath{\mathrm{SO}(#1)}}

\newcommand{\be}{\begin{equation}}
\newcommand{\ee}{\end{equation}}
\newcommand{\bea}{\begin{eqnarray}}
\newcommand{\eea}{\end{eqnarray}}

\newcommand{\neu}[1]{\ensuremath{\tilde{\chi}_{#1}^0}}

\newcommand{\chpm}[1]{\ensuremath{\tilde{\chi}_{#1}^{\pm}}}
\newcommand{\ttbar}{t \bar{t}}

\newcommand{\gsim}{\lower.7ex\hbox{$\;\stackrel{\textstyle>}{\sim}\;$}}
\newcommand{\lsim}{\lower.7ex\hbox{$\;\stackrel{\textstyle<}{\sim}\;$}}

\labelformat{subsection}{Section #1} 
\labelformat{equation}{Eq.~(#1)} 
\labelformat{figure}{Fig.~#1} 
\labelformat{subfigure}{Fig.~\thefigure#1} 
\labelformat{table}{Tab.~#1}

\def\be{\begin{equation}}
\def\ee{\end{equation}}
\def\bea{\begin{eqnarray}}
\def\eea{\end{eqnarray}}

\graphicspath{{./figures/}}

\begin{document}

\floatpagestyle{plain}

\pagenumbering{roman}

\renewcommand{\headrulewidth}{0pt}
\rhead{ OHSTPY-HEP-T-13-005 }
\fancyfoot{}

\title{\huge \bf{On the Viability of Thermal Well-Tempered Dark Matter in SUSY GUTs}}

\author[$\dag$]{Archana Anandakrishnan}
\author[$^\ddag$]{Kuver Sinha}

\affil[$\dag$]{\em Department of Physics, The Ohio State University,\newline
191 W.~Woodruff Ave, Columbus, OH 43210, USA  \enspace \medskip}

\affil[$^\ddag$]{\em Department of Physics, Syracuse University,\newline 
Syracuse, NY 13244, USA\enspace\enspace\enspace\enspace\enspace}

\maketitle
\thispagestyle{fancy}

\begin{abstract}\normalsize\parindent 0pt\parskip 5pt

In a scenario with heavy supersymmetric sfermions and decoupled supersymmetric Higgs sector, a well tempered neutralino is the remaining candidate for thermal single-component sub-TeV dark matter. Well tempered neutralinos are studied in the context of supersymmetric grand unified theories with third family Yukawa coupling unification. A global $\chi^2$ analysis is performed, including the observables $M_W,\ M_Z,\ G_F,\ \alpha_{em}^{-1},$  $\alpha_s(M_Z),\ M_t,\ m_b(m_b),\ M_\tau,\ b \rightarrow s \gamma,\ BR(B_s \rightarrow \mu^+ \mu^-)$, $M_{h}$ and $\Omega h^2$. Tensions in simultaneously fitting the Higgs and bottom quark masses while also avoiding gluino mass bounds from the LHC disfavors light Higgsinos with mass $\lsim 500$ GeV, ruling out light Bino/Higgsino dark matter candidates. Bino/Wino/Higgsino and Bino/Wino candidates fare somewhat better although they are fine-tuned and require departure from GUT scale gaugino mass universality (the example chosen here is the mixed modulus-anomaly pattern). Implications for dark matter direct detection of these models as well as collider signatures are briefly discussed. Independent of the thermal dark matter viability, these models will be severely constrained by the absence of a gluino at the next run of LHC.

\end{abstract}

\clearpage
\newpage

\pagenumbering{arabic}

\section{Introduction}

The major motivations for supersymmetric extensions of the Standard Model (SM) are gauge coupling unification, a solution to the hierarchy problem, and the existence (in $R$-parity conserving models) of a dark matter (DM) candidate. In particular, supersymmetry allows the gauge couplings to unify at an amazing precision of about $3\%$ \cite{Dimopoulos:1981yj, Dimopoulos:1981zb, Ibanez:1981yh, Sakai:1981gr, Einhorn:1981sx, Marciano:1981un}. Supersymmetric grand unified theories (SUSY GUTs) also allow for Yukawa coupling unification. For example, in minimal \SO{10} GUTS, each family of the standard model matter lives in one {\bf 16} dimensional representation, while the Higgses live in a {\bf 10} dimensional representation. The only renormalizable Yukawa coupling that can be written down in such theories is a $\lambda\ \bf{16\ 10\ 16}$, allowing for Yukawa coupling unification at the GUT scale. $t-b-\tau$ Yukawa unification requires $\tanb \simeq 50$ in order to reproduce the correct ratio between the top and the bottom quark masses. The constraints on the SUSY boundary conditions at the GUT scale coming from requiring Yukawa unification have been extensively studied \cite{Blazek:2001sb, Baer:2001yy, Blazek:2002ta, Tobe:2003bc, Auto:2003ys, Gogoladze:2011ce, Gogoladze:2011aa, Badziak:2011wm, Anandakrishnan:2012tj, Anandakrishnan:2013cwa, Ajaib:2013zha}.

In SUSY GUT models with universal gaugino masses at the GUT scale, the lightest supersymmetric particle (LSP) is typically a pure Bino, for which the DM relic density is too high in thermal histories\footnote{We note that non-thermal histories with a Bino LSP can satisfy the relic density 
if it is produced from the decay of a modulus with just the correct density and undergoes no further annihilation. This scenario, called the ``branching scenario", has been studied in \cite{Moroi:1999zb, Allahverdi:2012gk, Allahverdi:2013noa}.}. A pure thermal Bino could produce the observed relic density through coannihilation effects or the $A$-resonance, in a region where $M_A \sim 2 m_{\neu{1}}$, where $M_A$ is the mass of the pseudoscalar Higgs. Neither of these options are straightforward, however. Since each SM generation belongs to a {\bf 16}-dimensional representation there is one universal scalar mass for each family at the GUT scale. The third generation sfermion mass is constrained to be very heavy ($\gtrsim$ few TeV) due to flavor physics constraints from $b \rightarrow s \gamma$ transitions that are enhanced at large \tanb. Hence, coannihilation effects due to a light slepton that is nearly mass-degenerate with the LSP is difficult to obtain in these SUSY GUTs (assuming that the DM mass is itself in the sub-TeV range). Moreover, Yukawa unified GUTs prefer a large CP-odd Higgs mass to satisfy the $B_s \rightarrow \mu^+ \mu^-$ constraint which puts the $A$-resonance region also in tension. The popular remaining options are to opt for a candidate such as the pure Wino or Higgsino (either in a  non-thermal setting \cite{Allahverdi:2012wb} or as a part of multicomponent DM scenarios \cite{Baer:2008eq,Baer:2008yd}) or to rely on a (thermal) well-tempered candidate. We will not explore the non-thermal or multi-component options in this paper. Dark matter in Yukawa unifed SUSY GUTs have been discused in earlier works of Ref. \cite{Dermisek:2003vn, Dermisek:2005sw, Baer:2008jn, Baer:2008yd}.

The purpose of this paper is to study thermal, single-component well-tempered DM candidates \cite{ArkaniHamed:2006mb} satisfying the relic density in SUSY GUT models, given the constraints emerging from the Large Hadron Collider (LHC). For models with heavy scalars, the gluino is constrained by CMS and ATLAS searches \cite{ATLAS-CONF-2013-061,Chatrchyan:2013wxa} to be $M_{\tilde g} \gtrsim 1 - 1.2$ TeV, where the actual bound depends on the search strategy based on the final states from the decay of the gluino. On the other hand, the LHC has observed a new boson consistent with the SM Higgs with mass of about $125$ GeV \cite{:2012gu,:2012gk}. 

One of the first conclusions we draw from the current study is that light ($ \lsim \,\, \mathcal{O}(500)$ GeV) Higgsinos are difficult to obtain in Yukawa unified SUSY GUT models. This comes from simultaneously satisfying the observed Higgs mass and corrections to the bottom quark mass while also obtaining a gluino heavier than the LHC-excluded lower bound. For gluino $\gtrsim$ 1.2 TeV, we find tensions in obtaining a 125 GeV Higgs mass, in spite of the heavy scalars. The fact that light Higgsinos are difficult to obtain implies that light Bino/Higgsino ($\tilde{B}/\tilde{H}$) DM is disfavored in these models. This further implies that universal gaugino mass conditions, where $\tilde{B}/\tilde{H}$ is the only viable thermal well-tempered DM candidate (due to large mass separation between the Bino and the Wino), are similarly disfavored. 
 
We then proceed to explore Bino/Wino/Higgsino ($\tilde{B}/\tilde{H}/\tilde{W}$) and Bino/Wino DM ($\tilde{B}/\tilde{W}$). These cases require departure from gaugino mass universality at the GUT scale (to reduce the mass separation between Bino and Wino). The boundary condition we choose is mixed modulus-anomalous (mirage) mediation \cite{Choi:2007ka,Lowen:2008fm,Baer:2006tb} which is independently well-motivated from string constructions. We perform a global $\chi^2$ analysis including the observables $M_W,\ M_Z,\ G_F,\ \alpha_{em}^{-1},$ $\alpha_s(M_Z),\ M_t,\ m_b(m_b),\ M_\tau, $ $\ b \rightarrow s \gamma,\ BR(B_s \rightarrow \mu^+ \mu^-)$ and $M_{h}$. We then explore the best fit regions and study the thermal relic abundance, $\Omega h^2$.

Our main findings for $\tilde{B}/\tilde{H}/\tilde{W}$ and $\tilde{B}/\tilde{W}$ are the following: Compared to the case of universal gaugino masses and $\tilde{B}/\tilde{H}$ DM, these cases perform better but are also somewhat fine-tuned. A certain degree of fine-tuning is expected in general 
for Bino/Wino coannihilation effects to be operational. This is exacerbated by the additional constraints on the Higgsinos, outlined above and shown in detail in Section \ref{universalgaugino}. Nevertheless, we obtain regions of parameter space that have a well tempered neutralino DM and fit the other observables mentioned above at 90-95\% confidence level. 

The best fits are obtained for a certain range of $\alpha$ which is a parameter that controls the degree of departure from GUT-scale gaugino universality. $\alpha$ parametrizes the relative importance of (universal) modulus mediated and ($\beta$-function dependent) anomaly mediated contributions to the GUT scale gaugino masses. For $\alpha \, \sim \, 0$ (the limit in which anomaly mediated contributions vanish), the tensions associated with universal gaugino masses become manifest while for large $\alpha$ (the limit in which anomaly mediated contributions dominate), the well tempering is ruined and the LSP becomes Wino-like. In the intermediate regions of $\alpha$, the DM is a well tempered 
$\tilde{B}/\tilde{H}/\tilde{W}$ or $\tilde{B}/\tilde{W}$, where each component plays a role in the final annihilation cross section. Typically, in the case of $\tilde{B}/\tilde{H}/\tilde{W}$ DM, the Higgsino component in the LSP turns out to be less than $8 \%$, allowing us to (just) evade current direct detection limits although this type of DM will be detectable in the near future. The main annihilation channels in this region are $\neu{1} \neu{1} \rightarrow Zh$ and the coannihilation channels are $\neu{1} \neu{2}, \neu{1} \chpm{1} \rightarrow ZW$. In the case of $\tilde{B}/\tilde{W}$ DM, the Higgsino content is less than $1\%$ and the relic density is mainly satisfied through coannihilation effects while the spin independent scattering cross section is much below current direct detection bounds. 

It is important to note that independent of the thermal dark matter viability, these models will be severely constrained by the absence of a gluino at the next run of LHC. The best fit regions require a light gluino. The lower bound on the gluino mass is growing from the LHC data and the 14 TeV run at the LHC will conclusively test the regions discussed here. 

The plan of the paper is as follows: In Section \ref{yukawa}, we introduce Yukawa unified SUSY GUT models, discuss well-tempering in these scenarios, and also discuss the procedure we adopt for the obtaining the best fit points. In Section \ref{universalgaugino}, we present the case light Higgsinos and Bino/Higgsino DM in models with universal gaugino masses. Section \ref{BHdarkmattersection} contains our results for the $\tilde{B}/\tilde{H}/\tilde{W}$ and $\tilde{B}/\tilde{W}$ DM. Here, we first give a short background on the mirage boundary conditions and then present the spectra, fits, and properties of the well-tempered scenarios. We end with our conclusions. In the Appendices, we present analytical expressions for the annihilation cross section of $\neu{1} \neu{1} \rightarrow Zh$, which is one of the main annihilation channels of the $\tilde{B}/\tilde{H}/\tilde{W}$ case. We also show the best fit regions for several intermediate values of the parameter $\alpha$, which demonstrate that the fits get progressively better as one departs from gaugino universality.

\section{Model and Procedure}
\label{yukawa}

Supersymmetric parameters are heavily constrained when one requires Yukawa coupling unification. In the case of \tbtau\ Yukawa unification, one Yukawa coupling gives rise to the top, bottom and the tau masses. The only way to reproduce the hierarchy between the top and the bottom masses is by requiring large \tanb\  $\sim 50$. At large \tanb, the GUT scale SUSY parameters are further constrained. For example, in the large \tanb\ regime the down quark mass matrix and the CKM matrix elements receive significant corrections \cite{Blazek:1995nv}. Thus, requiring that the Yukawa couplings unify in addition to gauge coupling unification removes a lot of freedom from the GUT scale parameters. 

The model parameters are summarized in \ref{tab:parameters}. The model is defined by three gauge parameters, $\alpha_{G}, M_{G}, \epsilon_3$, where $\alpha_1(M_G) = \alpha_2(M_G) \equiv \alpha_G$, and  $\epsilon_3 = \frac{\alpha_3 - \alpha_G}{\alpha_G}$ is the GUT scale threshold corrections 
to the gauge couplings. There is one large Yukawa coupling, $\lambda$, such that, $\lambda_t(M_G) = \lambda_b(M_G) = \lambda_\tau(M_G) = \lambda$. Typically, there are small corrections to this relation at the GUT scale, coming from the off-diagonal Yukawa couplings to the first two families. 
Here we consider a third family model, since the SUSY spectrum (and relic abundance) does not heavily depend on these small off-diagonal Yukawa couplings. SUSY parameters defined at the GUT scale include a universal scalar mass for squarks and sleptons, $m_{16}$; universal gaugino mass, $M_{1/2}$; $m_{10}$, the universal Higgs mass; $A_0$, universal trilinear scalar coupling and $D$, the magnitude of Higgs splitting\footnote{Here we study the case of ``Just-so" Higgs splitting, D also fixes the magnitude of splitting for all scalar masses in the case of D-term splitting.}. Note that non-universal Higgs masses are necessary in order for radiative electroweak symmetry breaking in these models. We will also consider in general a parameter $\alpha$ that determines the ratio of anomaly mediation to gravity mediation contribution to the gaugino masses. The parameters $\mu, \ \tan\beta$ are obtained at the weak scale by consistent electroweak symmetry breaking. In the case of \tbtau\ unification, \tanb\ is restricted to be around 50\footnote{We also considered the possibility of relaxing \tbtau\ unification to \btau\ unification by allowing an independent Yukawa coupling for the bottom and tau, such that $\lambda_t = \lambda_u$ and $\lambda_b = \lambda_\tau = \lambda_d$. But, the fits that we obtained were consistent with \tbtau\ unification to within a few percent, and the extra parameter dependence was eliminated.}.

\begin{table}
\begin{center}
\renewcommand{\arraystretch}{1.2}
\scalebox{0.83}{
\begin{tabular}{|l||c|}
\hline
Sector &  Third Family Analysis  \\
\hline
gauge             & $\alpha_G$, $M_G$, $\epsilon_3$          \\
SUSY (GUT scale)  & $m_{16}$, $M_{1/2}$, $\alpha$, $A_0$, $m_{10}$, $D$        \\
textures          & $\lambda$                                         \\
SUSY (EW scale)   & $\tan \beta$, $\mu$                               \\
\hline
Total \#          &                                              12 \\
\hline
\end{tabular}
}
\caption{\footnotesize The model is defined by three gauge parameters, one large Yukawa coupling and 6 SUSY parameters defined at the GUT scale. The parameters $\mu, \ \tan\beta$ are obtained at the weak scale by consistent electroweak symmetry breaking.}
\label{tab:parameters}
\end{center}
\end{table}

\begin{table}
\begin{center}
\begin{tabular}{|l|l|l|c|}
\hline
\textbf{Observable} &  \textbf{Exp.~Value}   & \textbf{Ref.} & \textbf{Th. Error}   \\
\hline
\hline
$\alpha_3(M_Z)$          &  $0.1184\pm0.0007$               &
\cite{Beringer:1900zz}   &  0.5\% \\
$\alpha_{\text{em}}$  &  $1/137.035999074(44)$            &
\cite{Beringer:1900zz}   &  0.5\% \\
$G_\mu$             &  $1.16637876(7)\times10^{-5}\text{ GeV}^{-2}$ &
\cite{Beringer:1900zz}    & 1.0\% \\
$M_W$               &  $80.385\pm0.015\text{ GeV}$                     &
\cite{Beringer:1900zz}  & 0.5\% \\
$M_Z$              &   $91.1876 \pm 0.0021$                             &  
\cite{Beringer:1900zz}   &     0.5\%   \\
\hline
$M_t$               &  $173.5\pm1.0\text{ GeV}$                &
\cite{Beringer:1900zz}    &  0.5\%\\
$m_b(m_b)$               &  $4.18\pm0.03\text{ GeV}$ & \cite{Beringer:1900zz}   &
 0.5\% \\
$M_\tau$            &  $1776.82\pm0.16\text{ MeV}$                &
\cite{Beringer:1900zz}    & 0.5\% \\
\hline
$M_h$               &  $125.3\pm0.4\pm0.5\text{ GeV}$                &
\cite{:2012gu}    &  0.5\%\\
\hline
$\text{BR}(b \rightarrow s \gamma)$                &  $(343\pm21\pm7) \times
10^{-6}$      & \cite{hfag:2012-10-24} &   $(181 - 249) \times 10^{-6}$ \\
$\text{BR}(B_s  \rightarrow \mu^+ \mu^-) $          &  $(3.2\pm1.5)\times 10^{-9}$     
& \cite{:2012ct}   & $(1.5 - 4.7) \times 10^{-9}$  \\
\hline
$\Omega h^2$ & $0.1187\pm0.0017$ & \cite{Ade:2013ktc}  & 0.08 - 0.2 \\
\hline
\end{tabular}
\end{center}
\caption{\footnotesize The 12 observables that we compare in the SUSY GUT model and their experimental values. Capital letters denote pole masses. We take LHCb results into account, but use the average by Ref.~\cite{hfag:2012-10-24}. All experimental errors are $1\sigma$ unless otherwise indicated. Finally, the $Z$ mass is fit precisely via a separate $\chi^2$ function solely imposing electroweak symmetry breaking. $\Omega h^2$ is not included in the $\chi^2$ minimization procedure.
}
\label{tab:allobservables}
\end{table}

We follow the same procedure used in Ref.\cite{Anandakrishnan:2013cwa} to calculate 12 low energy observables. We use the 2-loop MSSM RGEs for both dimensionful and dimensionless parameters, integrate out the heavy scalars at their masses, evolve all parameters to the weak scale without the first two generation scalars. We use \maton \,\, to perform the RGE evolution, calculation of the couplings at the weak scale and the masses of the gauge bosons, top, bottom quarks and the $\tau$ lepton. For the calculation of Higgs mass, we define an effective theory at the scale $M_{\mathrm{SUSY}}$ and interface our calculation with the code by authors in Ref.\cite{Bernal:2007uv}. The flavor observables are calculated using the code \susyflavor \cite{Crivellin:2012jv},\, and the relic abundance is computed using \micromegas\cite{Belanger:2010pz}. A $\chi^2$ function is constructed by comparing the predicted observables, $y_i$ (except the relic abundance) with their measured values, $y_i^{\text{data}}$, given by the standard definition:
\begin{equation}
 \chi^2 = \sum_i \frac{|y_i - y_i^{\text{data}}|^2}{\sigma_i^2}
\end{equation}
where $\sigma_i$, is the assumed uncertainty in the calculation of the observable. The $\chi^2$ function is minimized to determine the best set of GUT scale parameters. This minimization procedure is carried out using MINUIT \cite{James:1975dr}. The input parameters are listed in \ref{tab:parameters}, and the observables and the theoretical errors assumed in estimating them are summarized in \ref{tab:allobservables}. Note that the model is defined in terms of 12 parameters of which, some are fixed during the minimization procedure. We compare our predictions to 11 observables, and calculate the 68\%, 90\%, and 95\% confidence level intervals using the $\chi^2$ function for the appropriate number of degrees of freedom (d.o.f.). 

The main constraints on the SUSY spectrum come from fitting the third family masses, the Higgs mass, and the flavor observables $b \rightarrow s \gamma$ and $B_s \rightarrow \mu^+ \mu^-$. Fitting the $t-b-\tau$ masses requires that \tanb\ be around 50, in order to reproduce the hierarchy between the top and the bottom masses, when $\lambda_t(M_G) = \lambda_b(M_G)$. In addition, fitting the bottom mass and the Higgs mass simultaneously requires light gluinos \cite{Anandakrishnan:2012tj}. The flavor physics observable $b \rightarrow s \gamma$ in MSSM is enhanced by a chargino-stop loop at large \tanb, and requires that the stops (and consequently all scalars) be heavier than a few TeV. $B_s \rightarrow \mu^+ \mu^-$ rate is also enhanced at large \tanb, and constrains CP odd Higgs mass, $M_A \gtrsim$ 1.2 TeV. This pushes the model to the decoupling regime ($M_A >> M_Z $), and the lightest Higgs is predicted to be SM-like.

\subsection{Well-tempering in SUSY Models}

The lightest neutralino in the MSSM is a mixture of the Bino ($\tilde{B}$), Wino ($\tilde{W}$), and Higgsino ($\tilde{H_1} ,\tilde{H_2}$) eigenstates 
\be
\neu{1} \, = \, N_{11} \tilde{B} + N_{12} \tilde{W} + N_{13} \tilde{H}_1 + N_{14} \tilde{H}_2 \,\,, 
\ee
where the $N_{1i}$ are the relevant weights along the different directions. 

There are essentially three options for obtaining the observed relic density: $(i)$ thermal single-component DM: the lightest neutralino is the sole DM candidate and the observed relic density is obtained by thermal freeze-out. $(ii)$ non-thermal single-component DM: the lightest neutralino 
is the sole DM candidate and has a non-thermal cosmological history. Both cases of thermal under-production and over-production can be accommodated in this case. $(iii)$ multi-component DM: the relic density is satisfied by the lightest neutralino along with one or more additional candidates, 
motivated by other physics (not necessarily supersymmetry). We will be interested in option $(i)$ above, reserving the study of non-thermal DM in these models for future work.

If $\neu{1}$ is a pure Higgsino, i.e. $N_{11} = N_{12} = 0$, the thermal relic density is approximately given by $\Omega h^2 \,  \sim \, 0.1 \left(\frac{\mu}{1 \, {\rm TeV}}\right)^2 \,\,.$ Clearly, TeV-scale Higgsinos are required to satisfy the thermal relic density. A pure Higgsino LSP with a thermal history is thus somewhat in conflict with naturalness arguments \footnote{Naturalness arguments generally make the prediction of a small $\mu$ term: since the mass of the $Z$ is set by the relation $m^2_{Z} = \mu^2 + m^2_{H_u}$ (in the large \tanb\ limit), small fine-tuning requires $\mu \, \sim \, m_Z$.}. Similarly, for a pure Wino, i.e. $N_{11} = N_{13} = N_{14} \sim  0$, the thermal relic density is satisfied for masses around $2.5$ TeV. A pure Bino over-produces DM in the scenario considered in this paper, with decoupled supersymmetric scalar and Higgs sectors. 

The correct relic density can be obtained if $\neu{1}$ is an appropriate admixture of Bino, Wino, and Higgsino states. There are three possibilities here: a $\tilde{B}/\tilde{H}$, $\tilde{B}/\tilde{W}$, and a $\tilde{B}/\tilde{H}/\tilde{W}$ DM candidate. The idea is that an over-abundant Bino acquires a larger Wino or Higgsino component, allowing it to annihilate rapidly to $Z,W,$ and $h$ final states. 
For the case of $\tilde{B}/\tilde{H}$ DM, the low-energy values of $M_1$ and $\mu$ are required to be close to each other, typically to within $10\%$. $\tilde{B}/\tilde{W}$ DM requires $M_1 \sim M_2$ and gives the correct relic density mainly through coannihilation of $\neu{1}$ with $\neu{2}$ and $\chpm{1}$.
 
\section{Light Higgsinos and $\tilde{B}/\tilde{H}$ DM with Universal Gaugino Masses} 
\label{universalgaugino}

In this section, we discuss the case of well tempered DM in Yukawa unified SUSY GUTs with universal gaugino masses. Since $\tilde{B}/\tilde{W}$ DM requires $M_1 \sim M_2$ at low energies, it cannot be obtained if one assumes gaugino unification at the GUT scale. The remaining option then is $\tilde{B}/\tilde{H}$ DM. This requires light Higgsinos. There is however, a very strong tension in obtaining $\tilde{B}/\tilde{H}$ DM in these models. This tension arises from an unusual candidate: the corrections to the bottom quark mass. 

It is well known that there are corrections to the bottom quark mass from a gluino-sbottom loop and a chargino-stop loop in MSSM that are $\mathcal{O}$(\tanb) enhanced. These large corrections can be written as 
\be \label{bottom1}
\frac{\delta m_b}{m_b} \simeq \frac{g_3^2}{12 \pi^2} \frac{\mu M_{\tilde{g}} \tanb}{m_{\tilde{b}}^2} + \frac{\lambda_t^2}{32 \pi^2} \frac{\mu A_t \tanb}{m_{\tilde{t}}^2} \,\, ,
\ee
where $m_{\tilde{b}}$, $m_{\tilde{t}}$ and $M_{\tilde{g}}$ are the sbottom, stop, and gluino masses respectively. When the stop and sbottom masses are roughly degenerate $\sim \tilde{m}$, we can rewrite the above expression as 
\be \label{bottom2}
\frac{\delta m_b}{m_b} \simeq \frac{\mu \tanb}{\tilde{m}^2} \left( \frac{g_3^2}{12 \pi^2} M_{\tilde{g}} + \frac{\lambda_t^2}{32 \pi^2} A_t \right)
\ee
%

\begin{figure}[ht!]
\begin{center}
\includegraphics[width=12cm]{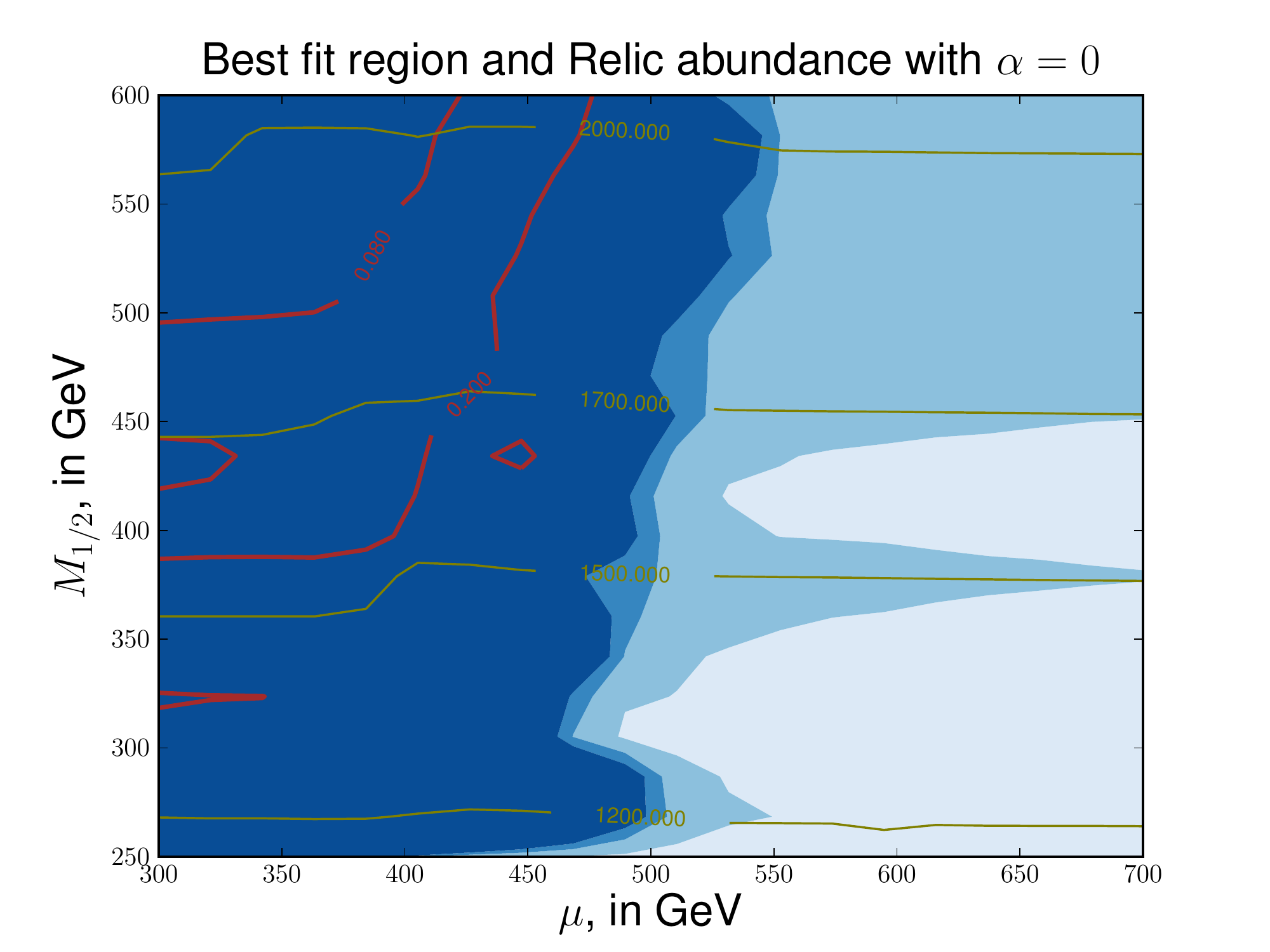}
\caption{\label{alpha0} {\footnotesize {\bf Universal Gaugino Masses and $\tilde{B}/\tilde{H}$ DM}: Best fit regions on a graph of $M_{1/2}$ versus $\mu$ in the case of $\alpha = 0$ (universal gaugino masses). The region between the red lines gives $\Omega h^2 = 0.08 \, - \, 0.2$, and the DM in this region is a well-tempered $\tilde{B}/\tilde{H}$. The olive curves are contours of constant gluino masses. The blue contour regions (lightest to darkest) represent the best fit regions at 68\%, 90\%, and 95\% CLs, respectively. Note that light Higgsinos (and hence light $\tilde{B}/\tilde{H}$ DM) are not preferred. }}
\end{center} 
\end{figure}


Light Higgsinos (small $\mu$) suppresses these corrections, which have to be at about a few \% (and negative), and are necessary to fit the bottom quark mass. This requirement places stringent constraints on the SUSY boundary conditions in the large \tanb\ regime. When the scalars are heavy, 
and when $\mu$ is small, one needs some fine-tuning between the gluino mass, $M_{\tilde{g}}$ and the $A_t$ parameter to generate these corrections. Collider constraints from LHC place significant lower bounds on the gluino mass, which is required to be heavier than $\sim 1000$ GeV in these models. Therefore, to get the correct $\sim$ few \% corrections, the trilinear coupling has to be large (and negative). When the absolute value of the trilinear coupling is forced to be larger than $\sim \sqrt{6} \tilde{m}$ (maximal mixing), it drives the Higgs mass to smaller values (or equivalently, the bad fit can be transferred to the bottom and tau mass). This tension between the bottom quark mass and the Higgs mass disfavors light Higgsinos in these models.

This tension is reinforced further when one considers well tempering in this context. If one does take the Higgsinos with mass $\sim 400$ GeV or greater, Binos are constrained to have similar mass, to give a well tempered DM. Due to gaugino universality, the gluinos then have mass $\sim 1300 - 2000$ GeV (but cannot be too much lighter), further forcing trilinear terms to go beyond maximal mixing. These tensions are reflected in \ref{alpha0} with universal gaugino masses at the GUT scale, where the the best fit points are displayed in the $M_{1/2}-\mu$ plane. The region between the red lines has relic density in the range $\Omega h^2 = 0.08 \, - \, 0.2$. The relic abundance is satisfied in this region due to a well tempering of Bino/Higgsino DM (due to universal conditions, the Bino-Wino mass splitting is too large to allow coannihilation). The olive curves are contours of constant gluino mass. In the plot, the regions under the blue contours (lightest to darkest) represent $\chi^2/d.o.f. = 1.2,\ 2.3,\
 3$ corresponding to 68\%, 90\%, and 95\% CLs, respectively. The regions were obtained by a parameter space scan with two degrees of freedom ($m_{16}, \mu,$ and $M_{1/2}$ were held fixed). Well tempering is obtained for comparable Bino and Higgsino masses, which gives gluinos in the mass range of $1600-2000$ GeV. The situation is worse for pure Higgsino DM which have to be heavier than a TeV and thus requiring even heavier gauginos. Therefore, in comparison with the phenomenological scenarios of CMSSM and NUHM, Yukawa unified GUTs are further constrained. Note that in the case of CMSSM and NUHM it has been pointed out that a TeV scale Higgsino LSP is preferred since it is unconstrained by current experiments and the correct Higgs mass can be easily obtained in this region \cite{Roszkowski:2009sm, Kowalska:2013hha}.  

We find from our $\chi^2$ analysis that in this case, the worst fits are to $m_b(m_b)$, and $M_\tau$, and $\alpha_s$ all of which have pulls $\gg 1$. It is also clear from \ref{alpha0} that regions with $\mu \lsim \mathcal{O}(500)$ GeV generally give large $\chi^2/d.o.f. > 3$ in these scenarios. It is abundantly clear that there is tension between requiring Yukawa unification and a well tempered DM, when one starts with universal gaugino masses. To retain the thermal single component DM scenario in these models, one must move away from the gaugino mass unification assumption at the GUT scale.  

\section{$\tilde{B}/\tilde{H}/\tilde{W}$ and $\tilde{B}/\tilde{W}$ DM} \label{BHdarkmattersection}

In this section, we relax gaugino mass universality at the GUT scale to add the possibility of having $\tilde{B}/\tilde{H}/\tilde{W}$ or $\tilde{B}/\tilde{W}$ DM candidates. A particularly well-motivated boundary condition is combining a universal (modulus) contribution with varying degrees of anomaly-mediated contributions. To this end, we first give some basic details about mixed modulus-anomaly or mirage mediation. We then give our main results for $\tilde{B}/\tilde{H}/\tilde{W}$ and $\tilde{B}/\tilde{W}$ DM.

\subsection{Effective Mirage Mediation}

Non-universal gaugino masses could arise when there is a non-singlet F-term under the GUT symmetry or when there are hybrid SUSY breaking mechanisms. States in non-singlet representations of the gauge symmetry could contribute large threshold corrections to the gauge couplings and could ruin precision gauge coupling unification. Hence we depart from gaugino universality by assuming that there is a hybrid SUSY breaking mechanism at the GUT scale like the ``mirage" mediation scenarios studied in string-inspired effective supergravity models \cite{Choi:2007ka,Lowen:2008fm, Baer:2006tb}. The boundary conditions considered here have been recently studied within Yukawa unified SUSY GUTs \cite{Anandakrishnan:2013cwa}. Non-universal gaugino mass scenarios were considered in the context of GUTs in the analyses of Refs.~\cite{Badziak:2011wm, Gogoladze:2011ce, Gogoladze:2011aa, Ajaib:2013zha, Guchait:2011hj}, where the SUSY breaking F-term that couples to the gaugino transforms as a non-singlet of the unified SO(10) gauge group, and hence gives rise to non-universal masses. 

The gaugino masses at the GUT scale obey a ``mirage" pattern:
\begin{equation} 
M_i = \left(1 + \frac{g_G^2 b_i \alpha}{16 \pi^2} \log \left(\frac{M_{Pl}}{m_{16}} \right) \right) M_{1/2} 
\label{mirage}
\end{equation} 
In the above expression, $\alpha$ is a parameter that controls the relative importance of the universal and anomalous contributions, and $b_i = (33/5, 1, -3) \; {\rm for} \; i = 1, 2, 3$, are the relevant $\beta$-function coefficients. We note that larger $\alpha$ leads to larger anomaly mediated contributions (and hence departure from the universal scenarios). The definition of $\alpha$ has appeared in different forms in the literature. The above expression matches with the definition in Ref.~\cite{Choi:2007ka} (with the assumption, however, that $m_{3/2} \simeq m_{16}$), but is related to the definition in Ref.~\cite{Lowen:2008fm} by the transformation $\frac{1}{\rho} = \frac{\alpha}{16 \pi^2} {\rm ln} \frac{M_{PL}}{m_{16}}$. For the scalars, we assume a universal mass and trilinear coupling at the GUT scale. Note that, while the gaugino and scalar soft terms can be obtained from the heterotic framework of Ref. \cite{Lowen:2008fm}, the issue of obtaining large A terms is still a model building challenge. Here, we let the phenomenology guide our choice of large trilinears at the GUT scale.  

 
\begin{figure}[h!]
\begin{center}
\includegraphics[width=\textwidth]{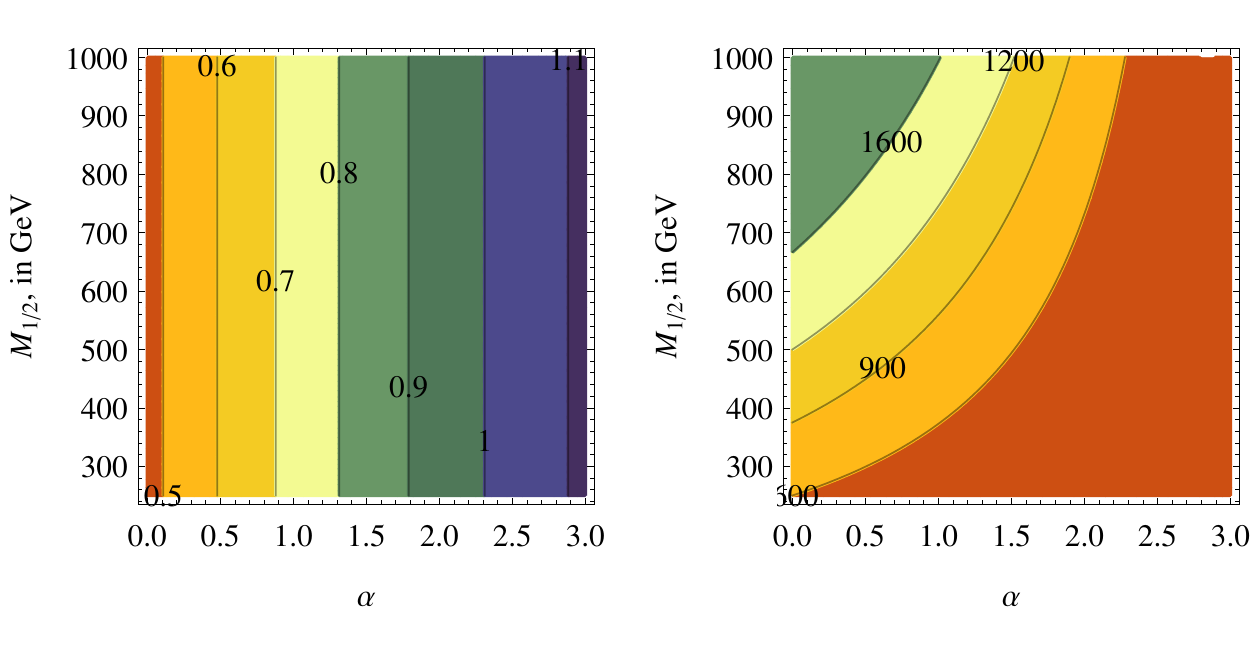}
\vspace{-1cm}
\caption{\label{m123} {\footnotesize The figure illustrates the relation between the gaugino masses at weak scale and the two GUT scale parameters, $M_{1/2}$ and $\alpha$ using simple tree level relations. {\it Left:} The ratio of the gaugino masses $M_1/M_2$ at the weak scale. {\it Right:} The gluino mass parameter $M_3$ at the weak scale.}}
\end{center}
\end{figure}


The additional degree of freedom $\alpha$, allows well-tempering by tuning the ratios of $M_1$ and $M_2$. To limit the regions of interest, we illustrate in \ref{m123}, the ratio of $M_1$ and $M_2$ and the value of $M_3$ obtained at the weak scale by a simple 1-loop analysis. We are interested in the regions where $|M_1| \leq |M_2|$, when the lightest neutralino starts to be a blend of Bino and Wino. This ratio of $|M_1|$ and $|M_2|$ is obtained when $\alpha$ is less than 3, below which the Wino becomes significantly lighter. \ref{m123} also shows that simultaneously satisfying the collider bound on the gluino mass further restricts $M_{1/2} > 250$ GeV. Note that the plot shows $M_3$ and there are additional corrections to the gluino pole mass. Once we identify the region of interest in the $M_{1/2} - \alpha$ parameter space we proceed to check if this region of the effective mirage mediation model is compatible with Yukawa unification. As an aside, we note that in Ref.~\cite{Anandakrishnan:2013cwa}, it was assumed that $\mu, \, M_{1/2} < 0$, and $\alpha \ge 4 $ such that $M_3 > 0, M_1, M_2 < 0$. This combination was useful to simultaneously satisfy the $b \rightarrow s \gamma$ constraint and the corrections to bottom quark mass. These boundary conditions led to very distinct spectrum and interesting phenomenology but as noted earlier, large values of $\alpha$ lead to anomaly mediation being the dominant source of SUSY breaking, and thus a pure Wino-like LSP. Therefore, they are outside the range of this work on thermal single component dark matter.

As $\alpha$ is gradually increased from $0$, there are two important effects that become apparent. The first one is that the Wino component of the LSP begins to increase. Therefore, the correct relic density is obtained for larger values of $\mu$ (compared to the universal case). Due to the constraints on light Higgsinos, this is preferable. On the other hand, since the beta-function coefficient is negative for SU(3), the gluino mass decreases with increasing $\alpha$. Then the model starts conflicting with the LHC results as $\alpha$ increases. Below are our results for the 
case of $\alpha = 1.5$ in \ref{mirage}. In Appendix \ref{appotheralpha}, we discuss the intermediate values of $\alpha = 0.5$ and $\alpha = 1$, where we find progressively improving (compared to the $\alpha = 0$ case) results for the $\chi^2$ fit. There, we also show that the model starts 
to conflict severely with the LHC results at around $\alpha \sim 2$. There is a finite volume of parameter space that permits well-tempering and the entire volume is within the reach of the LHC and DM experiments.  

\subsection{Results for $\tilde{B}/\tilde{H}/\tilde{W}$ DM}


\begin{figure}[ht!]
\begin{center}
\includegraphics[width=12cm]{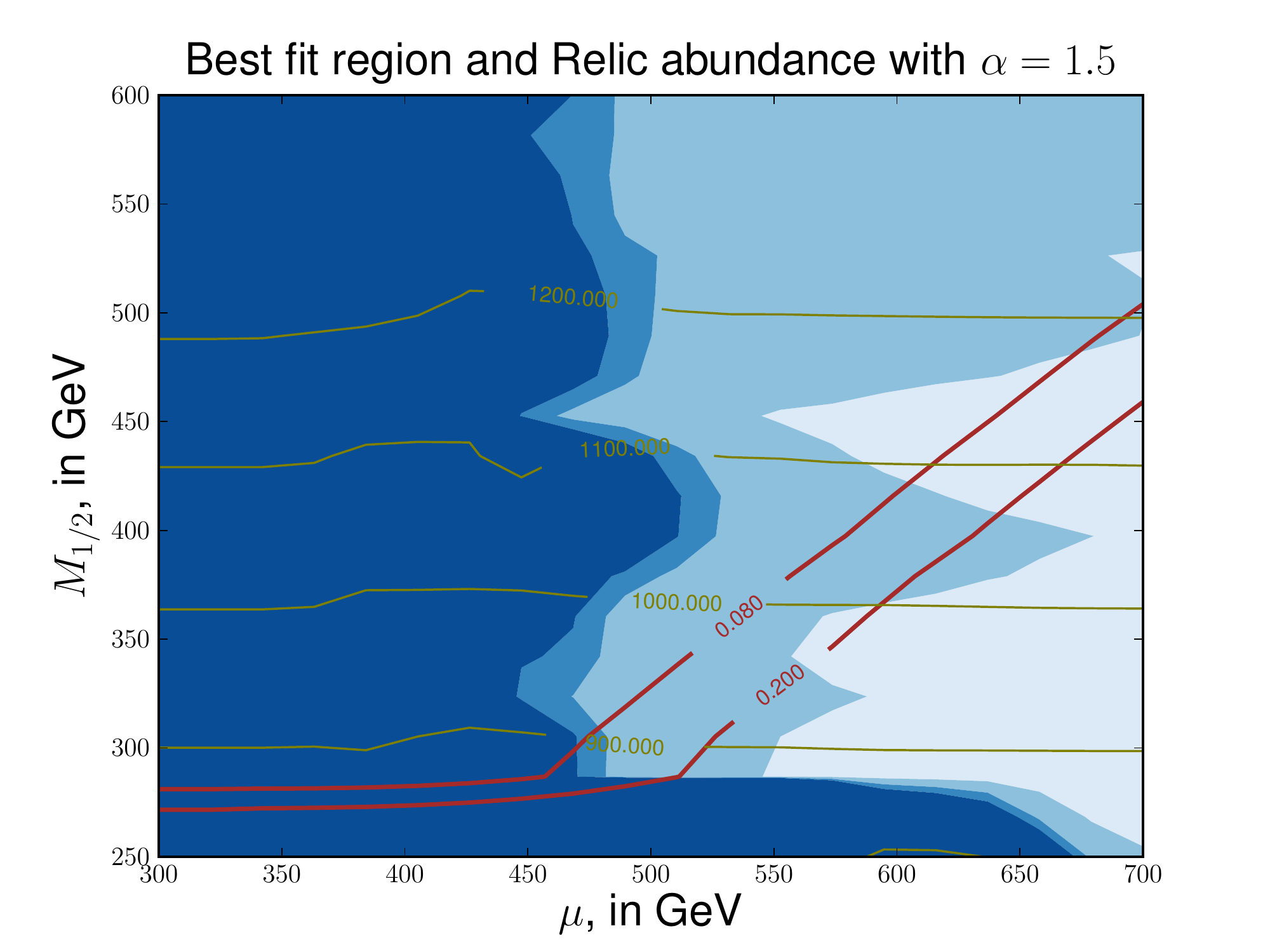}
\caption{\label{alpha1.5} {\footnotesize {\bf $\tilde{B}/\tilde{H}/\tilde{W}$ DM}: Best fit regions on a graph of $M_{1/2}$ versus $\mu$ in the case of $\alpha = 1.5$. The region between the red lines gives $\Omega h^2 = 0.08 \, - \, 0.2$, and the LSP in this region is a Bino/Wino/Higgsino admixture. The olive contours of constant gluino masses show that the gluino is lighter compared to the universal gaugino mass scenario. The blue contour regions (lightest to darkest) represent the best fit regions at 68\%, 90\%, and 95\% CLs, respectively.}}
\end{center} 
\end{figure}


The gaugino masses are maximally split in the universal case. As anomaly contributions increase with increasing $\alpha$, the mass splitting between the gaugino masses decrease. Due to the greater proximity of Bino and Wino masses, now, in fact, $\tilde{B}/\tilde{H}/\tilde{W}$ and $\tilde{B}/\tilde{W}$ DM can both be options for the well tempered DM candidate. Moreover, for the same Bino mass, the gluino can now be lighter ($\gsim 1100$ GeV), reducing the tension with fitting 
the $b$ mass. Indeed, we find now that we find better fits to all the observables in \ref{tab:allobservables}. This is reflected in \ref{alpha1.5} where we present fits with $\alpha = 1.5$. The region between the blue and red lines gives $\Omega h^2 = 0.08 \, - \, 0.2$. We see that this region now extends to areas of smaller $\chi^2$, with gluino mass $\sim 1100$ GeV. As in the universal case $\alpha = 0$, the region with low $\mu \lsim 450$ GeV gives a large value of $\chi^2$ and this region is qualitatively similar to the universal case. However, with larger $\mu \gsim 600$ GeV a region of low $\chi^2$ opens up and the relic density band extends into it. 

We present a sample benchmark point in \ref{benchmarkspectrumbhw}. The best fit point obtained here is very similar to the benchmark points discussed in Dermisek-Raby model with universal gaugino masses \cite{Anandakrishnan:2012tj}. The SUSY boundary conditions at the GUT scale are typically:
 \begin{align}
m_{16} > \text{few TeV}; &\qquad& m_{10} \sim \sqrt{2} m_{16}; \nonumber \\
A_0 \sim -2 m_{16}; &\qquad& \mu, M_{1/2} << m_{16}; \nonumber \\
&\qquad \qquad \tanb \sim 50&   \nonumber
\end{align}
The first two family of scalars are very heavy ($\sim$ 20 TeV), and the spectrum has the inverted scalar mass hierarchy with the third family of squarks and sleptons between $3-6$ TeV. These scalars are out of the reach of the LHC. The gauginos are light and the LSP (the DM candidate) is a well tempered $\tilde{B}/\tilde{H}/\tilde{W}$ mixture. The gaugino sector differs from the previous studies, and we are able to satisfy the relic abundance. The gluino tends to be lighter than the universal case due to the presence of the anomaly contribution. We have increased the overall scale 
of the gauginos, and the LSP mass but the gluino remains light enough to contribute the required corrections to the bottom quark mass. Note that all the extra Higgses of SUSY are heavy and thus the lightest Higgs is purely SM like. Finally, the gluino decays into a variety of final states. The final states should include many jets, b-jets and a large amount of missing energy, and large total transverse jet momentum. We expect the bounds from present LHC results to be similar to the model discussed in \cite{Anandakrishnan:2013nca} and we expect the 20 fb$^{-1}$ data from 8 TeV LHC to rule out gluinos lighter than 1 TeV.  

\vspace{1cm}
\begin{table}[ht!]
\begin{center}
\scalebox{0.9}{
\renewcommand{\arraystretch}{1.2}
\begin{tabular}{|l|l r |l r |l r |l r|}
\hline
GUT scale parameters & $m_{16}$ & 20000 & $M_{1/2}$ & 450  & $A_0$  & -40461 & $\alpha$ & 1.5 \\
 & $m_{H_d}$ & 27495 & & & $m_{H_u}$ & 23748 & & \\
& 1/$\alpha_G$ & 26.17  & $M_G$ & 2.13 $\times 10^{16}$ & $\epsilon_3$ & 0  & $\lambda$ & 0.59  \\
\hline
 EW parameters & $\mu$ & 660 & & & tan$\beta$ & 49.75 &  & \\
\hline
Fit & Total $\chi^2$& 1.84 & & & & & & \\
\hline
Spectrum &  $m_{\tilde u}$ & $\sim$ 20000  & $m_{\tilde d}$  & $\sim$ 20000   & $m_{\tilde e}$  & $\sim$20000 & &  \\
 & $m_{\tilde t_1}$ & 3612  & $m_{\tilde b_1}$  & 5053   & $m_{\tilde \tau_1}$  & 6867 & $M_{\tilde g}$ & 1130  \\
& $m_{\tilde\chi^0_1}$   & 474.5  & $m_{\tilde\chi^0_2}$  &  556.7  & $m_{\tilde\chi^0_3}$   & 693.6  & $m_{\tilde\chi^0_4}$ & 662.9\\
 & $m_{\tilde\chi^+_1}$   & 554.9  &  &  & $m_{\tilde\chi^+_1}$   & 691.5 & & \\
& $M_A$   & 1915.3  & $M_H^{\pm}$  & 1916.9 & $M_H$   & 1932.5 & $M_h$ & 121  \\
 \hline
DM & $\Omega h^2$& 0.121 & & & & & & \\
\hline
Gluino Branching Fractions & $ tb\widetilde{\chi}^\pm_2$ & 19\% & $ tb\widetilde{\chi}^\pm_1$ & 17\% & $ g \widetilde{\chi}^0_4$ & 17\% & $ t\bar{t}\widetilde{\chi}^0_4 $& 13\% \\
  & 
$ g \widetilde{\chi}^0_3$ & 12\% & $ t\bar{t}\widetilde{\chi}^0_2 $ & 9\% & $ t\bar{t}\widetilde{\chi}^0_3 $& 5\%& 
$ g \widetilde{\chi}^0_2$ & 4\% \\
\hline
\end{tabular}}
  \caption{\footnotesize Spectrum at the benchmark point for $\tilde{B}/\tilde{H}/\tilde{W}$ DM.}
  \label{benchmarkspectrumbhw} 
\end{center}
\end{table}

\subsection{ $\tilde{B}/\tilde{H}/\tilde{W}$ DM Annihilation and Scattering Cross Sections}

\begin{figure}[!ht]
\begin{center}
\includegraphics[width=8cm]{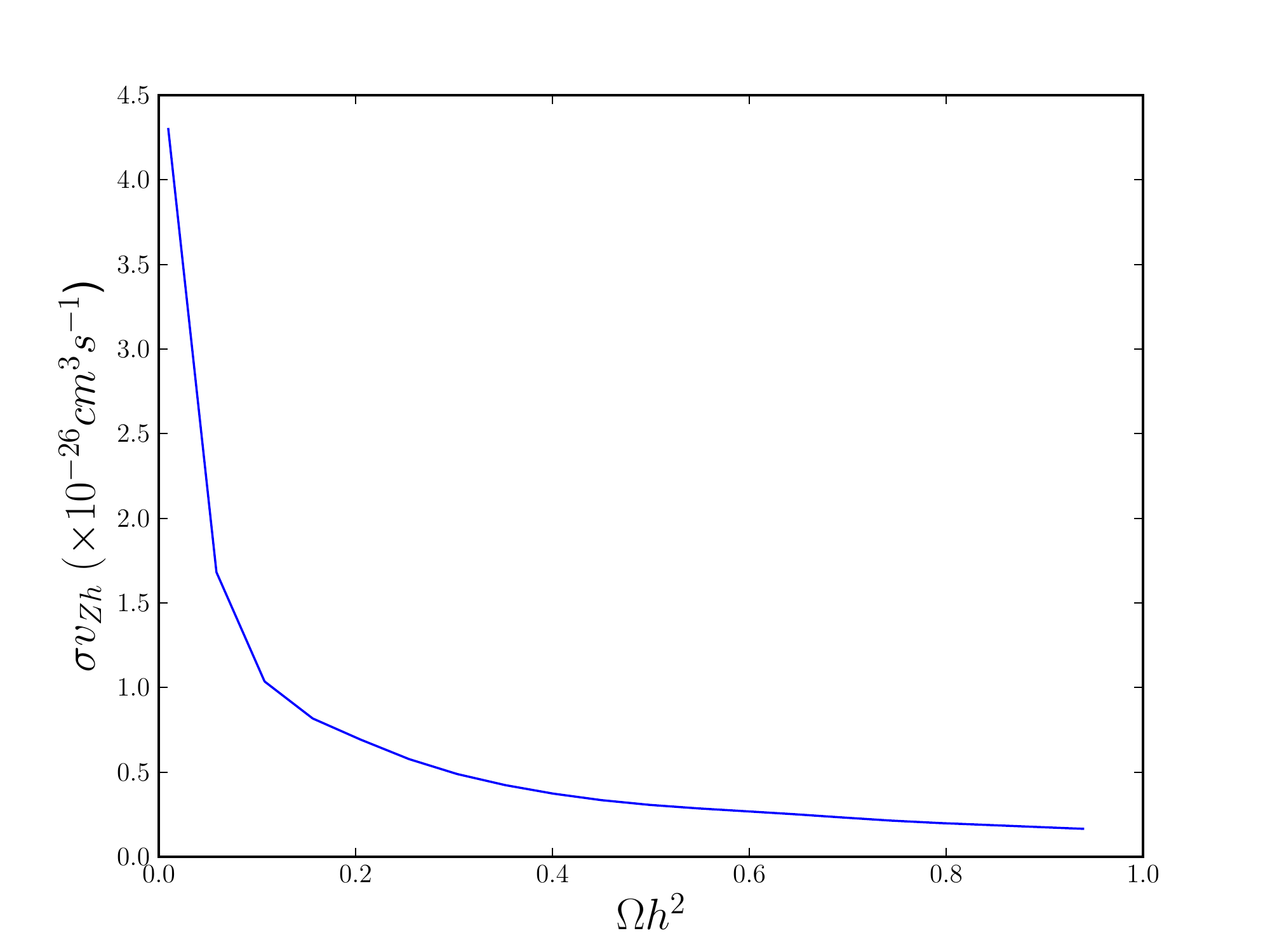}
\includegraphics[width=8cm]{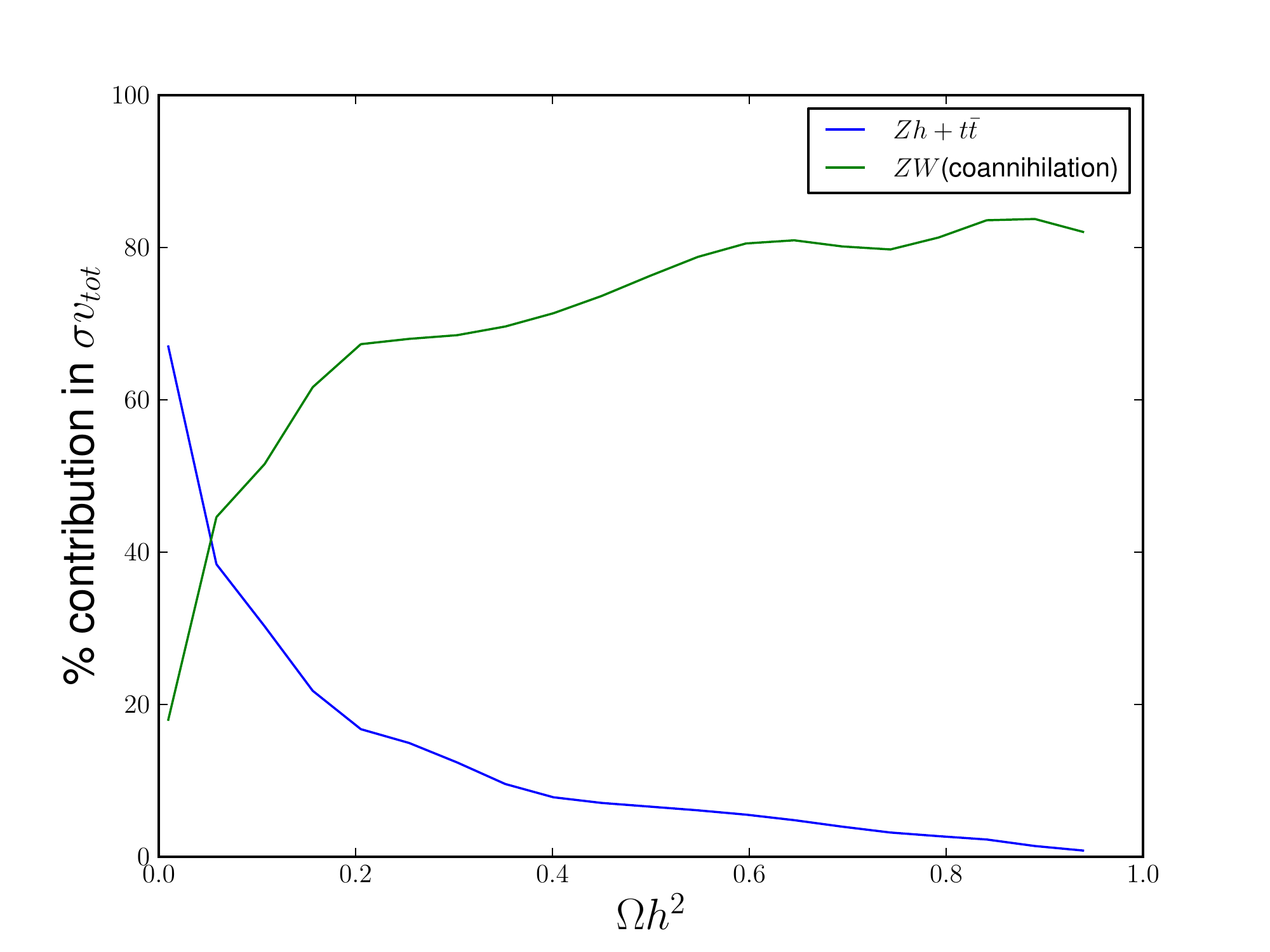}
\caption{\label{Zh_cross_section} {\footnotesize {\bf Left panel}: The annihilation cross-section $\neu{1} \neu{1} \rightarrow Zh$ as a function of the relic density for $\alpha = 1.5$. The cross-section is large for large Bino/Higgsino mixing. As the Higgsino becomes decoupled, the cross-section for this annihilation channel becomes negligible. For a theoretical computation, we refer to Appendix \ref{appZhcalculation}. {\bf Right panel}: The relative importance of the $Zh + \ttbar$ channel and the $ZW$ channel as a function of the relic density. For low relic density, the coannihilation channels $\neu{1} \neu{2}, \neu{1} \chpm{1} \rightarrow  ZW $ dominate. The DM is then a Bino/Wino well tempered neutralino. For larger separation between Bino and Wino, the coannihilation becomes ineffective and the relic density increases as $\neu{1}$ becomes predominantly Bino. The main annihilation channel then becomes  $\neu{1} \neu{1} \rightarrow  \ttbar $. For $\Omega h^2 \sim 0.1$, the annihilation to $Zh$ and $ZW$ are comparable.}}
\end{center} 
\end{figure}

The main annihilation channels of $\neu{1}$ in this region are displayed in \ref{Zh_cross_section}. 

The left panel shows the annihilation cross section $\neu{1} \neu{1} \rightarrow Zh$ as a function of the relic density. The cross section is enhanced for large Bino/Higgsino mixing. As the Higgsino becomes decoupled, the cross section for this annihilation channel becomes negligible. These results have been obtained using \micromegas\ and we have verified the result theoretically; the main formulae are collected in Appendix \ref{appZhcalculation}. The right panel shows the relative importance of the $Zh + \ttbar$ channel and the $ZW$ channel as a function of the relic density. For low relic density, the coannihilation channels $\neu{1} \neu{2}, \neu{1} \chpm{1} \rightarrow  ZW $ dominate. The DM is 
then a Bino/Wino well tempered neutralino. For larger separation between Bino and Wino, the coannihilation becomes ineffective and the relic density increases as $\neu{1}$ becomes predominantly Bino. The main annihilation channel then becomes  $\neu{1} \neu{1} \rightarrow  \ttbar $. We note that at the sweet spot $\Omega h^2 \sim 0.1$, the annihilation to $Zh$ and coannihilation to $ZW$ are comparable, indicating that the DM is a well tempered Bino/Wino/Higgsino type. 

\begin{table}[h!]
\begin{center}
\begin{tabular}{|l|l r|}
\hline
Annihilation cross section & $\neu{1}\neu{1} \rightarrow Zh$ & $52\%$ \\
 &$\neu{1}\chpm{1} \rightarrow ZW$ & $24\%$ \\
& $\neu{1}\neu{1} \rightarrow ZZ$  & $11\%$  \\
& $\neu{1}\neu{1} \rightarrow \ttbar $  & $6\%$  \\
\hline
 Scattering cross section (pb) & $1.6 \times 10^{-8}$ &  \\
\hline
\end{tabular}
  \caption{Annihilation and spin-independent scattering cross sections at the benchmark point with $\tilde{B}/\tilde{H}/\tilde{W}$ DM. Only channels contributing greater than $1\%$ to the total annihilation cross section are shown.}
  \label{darkmatterannihilation}
\end{center}
\end{table}

In \ref{darkmatterannihilation}, we display the annihilation channels and scattering cross section of the DM candidate corresponding to the benchmark point in \ref{benchmarkspectrumbhw}. The main annihilation channel is $\neu{1} \neu{1} \rightarrow Zh$ with a value of $\langle \sigma v \rangle_{Zh} = 1.02 \times 10^{-26}$ cm$^{3}$s$^{-1}$. For the benchmark point, the spin independent scattering cross-section is $1.6 \times 10^{-8}$ pb. We note that the scattering cross-section for this $m_{\neu{1}}$ is just at the exclusion limit at $90\%$ CL from XENON100 \cite{Aprile:2012nq}. Direct detection bounds for well tempered scenarios are getting particularly stringent. For $\mu > 0$, XENON100  exclusion limits force $m_{\neu{1}} \, \gsim \, 600$ GeV in the large \tanb \,\, limit \cite{Cheung:2012qy}. Typically, Bino/Higgsino mixing larger than $10 \%$ starts to conflict with exclusion bounds from XENON100 in these regions \cite{Dutta:2013sta, Han:2013gba} (for our benchmark scenario, the degree of Bino/Higgsino mixing is $\sim 8\%$). We thus see that the $\tilde{B}/\tilde{H}/\tilde{W}$ well-tempering is particularly interesting since it is likely to be conclusively probed by XENON1T \cite{Aprile:2012zx}. In fact, by changing $\alpha$, one can control the degree to which Bino/Wino coannihilation contributes to the relic density, and also the scattering 
 cross section. 

\subsection{Results for $\tilde{B}/\tilde{W}$ DM} 

\begin{table}[h!]
\begin{center}
 \scalebox{0.9}{
\renewcommand{\arraystretch}{1.2}
\begin{tabular}{|l|l r |l r |l r |l r|}
\hline
GUT scale parameters & $m_{16}$ & 29781 & $M_{1/2}$ & 600  & $A_0$  & -60395 & $\alpha$ & 2.3 \\
  & $m_{H_d}$ & 40724 & & & $m_{H_u}$ & 35237 & & \\
& 1/$\alpha_G$ & 26.35  & $M_G$ & 2.36 $\times 10^{16}$ & $\epsilon_3$ & 0  & $\lambda$ & 0.59  \\
 \hline
 EW parameters & $\mu$ &1200 & & & tan$\beta$ & 49.65 &  & \\
\hline
Fit & Total $\chi^2$& 1.72 & & & & & & \\
\hline
Spectrum &  $m_{\tilde u}$ & $\sim$ 29337  & $m_{\tilde d}$  & $\sim$ 29337   & $m_{\tilde e}$  & $\sim$29498 & &  \\
  & $m_{\tilde t_1}$ & 5832  & $m_{\tilde b_1}$  & 8078   & $m_{\tilde \tau_1}$  &10565 & $M_{\tilde g}$ & 1135  \\
& $m_{\tilde\chi^0_1}$   & 799.0  & $m_{\tilde\chi^0_2}$  &  835.9  & $m_{\tilde\chi^0_3}$   & 1201  & $m_{\tilde\chi^0_4}$ & 1210\\
  & $m_{\tilde\chi^+_1}$   & 835.5  &  &  & $m_{\tilde\chi^+_1}$   & 1209 & & \\
& $M_A$   & 3093  & $M_H^{\pm}$  & 3094 & $M_H$   & 3131 & $M_h$ & 123  \\
  \hline
DM & $\Omega h^2$& 0.099 & & & & & & \\
\hline
Gluino Branching Fractions & $ g \widetilde{\chi}^0_2$ &  55\% & $ g \widetilde{\chi}^0_1$ &   31\% &  $ tb\widetilde{\chi}^\pm_1$&   12\% 
& &    \\
\hline
\end{tabular}}
   \caption{\footnotesize  Spectrum at the benchmark point with $\tilde{B}/\tilde{W}$ DM. }
  \label{benchmarkspectrum2} 
\end{center}
\end{table}

As $\alpha$ increases further, the wino starts becoming a substantial component of the LSP and the relic abundance is satisfied for larger values of $\mu$. In this region we obtain a $\tilde{B}/\tilde{W}$ DM. A typical spectrum is shown in \ref{benchmarkspectrum2}. Note that we have chosen a higher scale for the supersymmetric scalar masses, $m_{16}$, so as to obtain a better fit to the Higgs mass. This also gives rise to larger corrections to the gluino mass. The gluino is $\gtrsim $ 1100 GeV, in spite of $\alpha > 2$, in comparison with \ref{otheralpha}, where $m_{16}$ was chosen to be 20 TeV. The other qualitative features of the spectrum remain unchanged. The main channels as well as the spin independent scattering cross section are shown in \ref{darkmatterannihilation2}. The annihilation cross section of the DM is dominated by various coannihilation processes among $\neu{1}, \neu{2},$ and $\chpm{1}$. The spin independent scattering cross section is well below XENON100 limits, as expected in this case.

\begin{table}[h!]
\begin{center}
\begin{tabular}{|l|l r|}
\hline
Annihilation cross section & $\neu{1}\neu{1} \rightarrow Zh$ & $6\%$ \\
 &$\neu{1}\chpm{1} \rightarrow ZW, Wh$ & $24\%$ \\
& $\neu{2}\chpm{1} \rightarrow ff, ZW$  & $32\%$  \\
& $\neu{2}\neu{2} \rightarrow WW $  & $4\%$  \\
& $\chpm{1}\chpm{1} \rightarrow WW, ZZ, ff $  & $21\%$  \\
\hline
 SI Scattering cross section (pb) & $3.5 \times 10^{-9}$    &  \\
\hline
\end{tabular}
  \caption{Annihilation and spin-independent scattering cross sections at the benchmark point with $\tilde{B}/\tilde{W}$ DM. Only channels contributing greater than $1\%$ to the total annihilation cross section are shown.}
  \label{darkmatterannihilation2}
\end{center}
\end{table}

\section{Comments and Summary}
In this work, we have investigated the interplay between four physics components within the context of Yukawa unified supersymmetric GUTS. 

\begin{itemize}
\item Thermal Relic Abundance.
\item Low energy observables including fermion masses and flavor. 
\item 125 GeV Higgs mass.
\item Collider bounds on the gluinos. 
\end{itemize}

We have studied the viability of a thermal DM candidate in Yukawa unified SUSY GUTS. Given the model parameters at the GUT scale, we have used the low energy observables $M_W,\ M_Z,$ $\ G_F,$ $\alpha_{em}^{-1},$ $\alpha_s(M_Z),\ M_t,\ m_b(m_b),\ M_\tau,\ b \rightarrow s \gamma,\ 
BR(B_s \rightarrow \mu^+ \mu^-)$ and $M_{h}$ to constrain the parameter space. It is well known that uniformly heavy scalars and decoupled SUSY Higgs sector are required in these models to evade flavor physics constraints from $b \rightarrow s \gamma,\ BR(B_s \rightarrow \mu^+ \mu^-)$. The heavy scalars make it impossible to achieve coanhhilation scenarios with staus or the CP-odd Higgs. The model also prefers a light gaugino spectrum and therefore disfavors Higgsino LSP. Then, the remaining option to obtain a thermal DM candidate in these models is to consider a well-tempered neutralino that is either $\tilde{B}/\tilde{H}$ or $\tilde{B}/\tilde{W}$ or $\tilde{B}/\tilde{H}/\tilde{W}$ admixture.

The first observation is that light Higgsinos with mass $\lsim \mathcal{O}(500)$ GeV are difficult to obtain in these models. In particular, the tension arises from simultaneously obtaining acceptable corrections to the bottom quark mass and the mass of the Higgs, as well as evading the lower bound on the mass of the gluino mass. This is clearly represented in \ref{alpha0}, where Higgsino masses less than $\sim \mathcal{O}(500)$ GeV uniformly have a large $\chi^2/d.o.f.$ for gluino mass $\gsim 1100$ GeV. This makes light well-tempered $\tilde{B}/\tilde{H}$ DM less preferred; and hence, universal gaugino masses at the GUT scale also less preferred, since $\tilde{B}/\tilde{H}$ DM is the only well-tempered option with that boundary condition.

We studied the cases of $\tilde{B}/\tilde{H}/\tilde{W}$ and $\tilde{B}/\tilde{W}$ DM by allowing for non-universal gaugino masses at the GUT scale, in a mirage pattern. Non-universality of the gauginos at the GUT scale is required to compress the mass splitting between the Bino and the Wino. We find that there are small regions of parameter space that can accommodate all the low energy observables including a 125 GeV Higgs that do not conflict with the LHC constraints on the gluino. In both cases there is a considerable degree of fine-tuning between the Higgsino, third generation squark, and gluino masses in order to satisfy all the constraints. 

The best fit regions require a light gluino. The lower bound on the gluino mass is growing from the LHC data and the 14 TeV run at the LHC will conclusively test the regions discussed here. For gluino $\sim$ 1.3 TeV, we find tensions in obtaining a 125 GeV Higgs mass, in spite of the heavy scalars in the model. Our findings are that the trilinear coupling $A_t$ is pushed beyond maximal mixing, and hence drives the Higgs mass to smaller values. On the other hand, the $\tilde{B}/\tilde{H}/\tilde{W}$ will also be probed by XENON1T. While there are still small pockets of the parameter space with accidental degeneracies, we find that these are considerably fine-tuned and may even require DM with mass larger than a TeV. It is important to note that independent of the thermal dark matter viability, these models will be severely constrained by the absence of a gluino at the next run of the LHC. Hints of the gluino at the next run of the LHC without any immediate detection from direct DM searches would leave only the $\tilde{B}/\tilde{W}$ as a viable scenario.

\section{Acknowledgements}

 We would like to thank Stuart Raby for helpful discussions, and Radovan Dermisek for the code \maton\ which has been adapted to do a major portion of the top-down analysis. We also acknowledge discussions with Genevieve Belanger, Alexander Pukhov, and Paolo Gondolo. K.S. is supported by NASA Astrophysics Theory Grant NNH12ZDA001N. A.A. is partially supported by DOE grant DOE/ER/01545-901.

\begin{appendices}


\section{Annihilation channel $\neu{1} \neu{1} \rightarrow Zh$} \label{appZhcalculation}

In this section, we present the analytical expression for the annihilation cross section of DM into $Zh$ final states, following \cite{Kamionkowski:1991nj}. The annihilation of DM into $Zh$ final state occurs due to two contributions: $s$-channel exchange of a $Z$ and $t$- and $u$- channel exchange of all four neutralinos. The cross section in the limit of $v_{\rm rel} \rightarrow 0$ is given by
\be
\sigma_{Zh} v_{\rm rel} \, = \, \frac{k X_{Zh}}{32 \pi m^3_{\neu{1}}} \,\, ,
\ee
where
\be
k = \left[m^2_{\neu{1}} - \frac{1}{2} (m^2_Z + m^2_h) + \frac{(m^2_Z - m^2_h)^2}{16 m^2_{\neu{1}}}   \right]^{1/2}
\ee
is the momentum of the outgoing particles. In the limit of heavy Higgs partners, the quantity $X$ is given by
\be \label{XinZh}
X_{Zh} = 2k^2 \frac{m^2_{\neu{1}}}{m^2_Z} \left[ \frac{z F_{nn}}{m^2_Z} + 
\sum_K \frac{2gM_{2nk} F_{nk} (m_{\neu{k}} - m_{\neu{1}})}{t - m^2_{\neu{k}}}  \right] ^2 \,\,.
\ee
The various quantities in \ref{XinZh} are: $(i)$ the coupling at the $hZZ$ vertex $z = g m_Z \sin{\beta - \alpha} / \cos {\theta_W}$ ($g$ is the $SU(2)_L$ coupling constant), $(ii)$ the coupling of the $Z \neu{i} \neu{j}$ vertex $F_{ij} = g (Z_{i3}Z_{j3} - Z_{i4}Z_{j4})/2\cos{\theta_W}$, $(iii)$ the coupling at the $H^0_i \neu{j} \neu{k}$ vertex $M_{ijk}$ available in \cite{Griest:1989zh}, and $(iv)$ $t = \frac{m^2_Z - m^2_h}{2} - m^2_{\neu{1}}$. The sum is over all neutralinos. 

\section{$\alpha = 0.5,\ 1\ \&\ 2$}
\label{appotheralpha}

In the main text, we have given the examples of $\alpha = 0$ (universal gaugino masses) and $\alpha = 1.5$. The problems with the universal gaugino mass case were described, and the eventual benchmark at $\alpha = 1.5$ was presented. In this Appendix, we give the results for several intermediate values of $\alpha$. We note that the region with relic densty $\Omega h^2 = 0.08 \, - \, 0.2$ starts extending more and more into the region with small $\chi^2$, as the $b$-mass fitting improves due to the reasons mentioned in the text. For $\alpha = 2.0$, the $\chi^2$ fit is the best; however, the trade off is that the gluino becomes too light. 

\begin{figure}[!htp]
\begin{center}
\includegraphics[width=8cm]{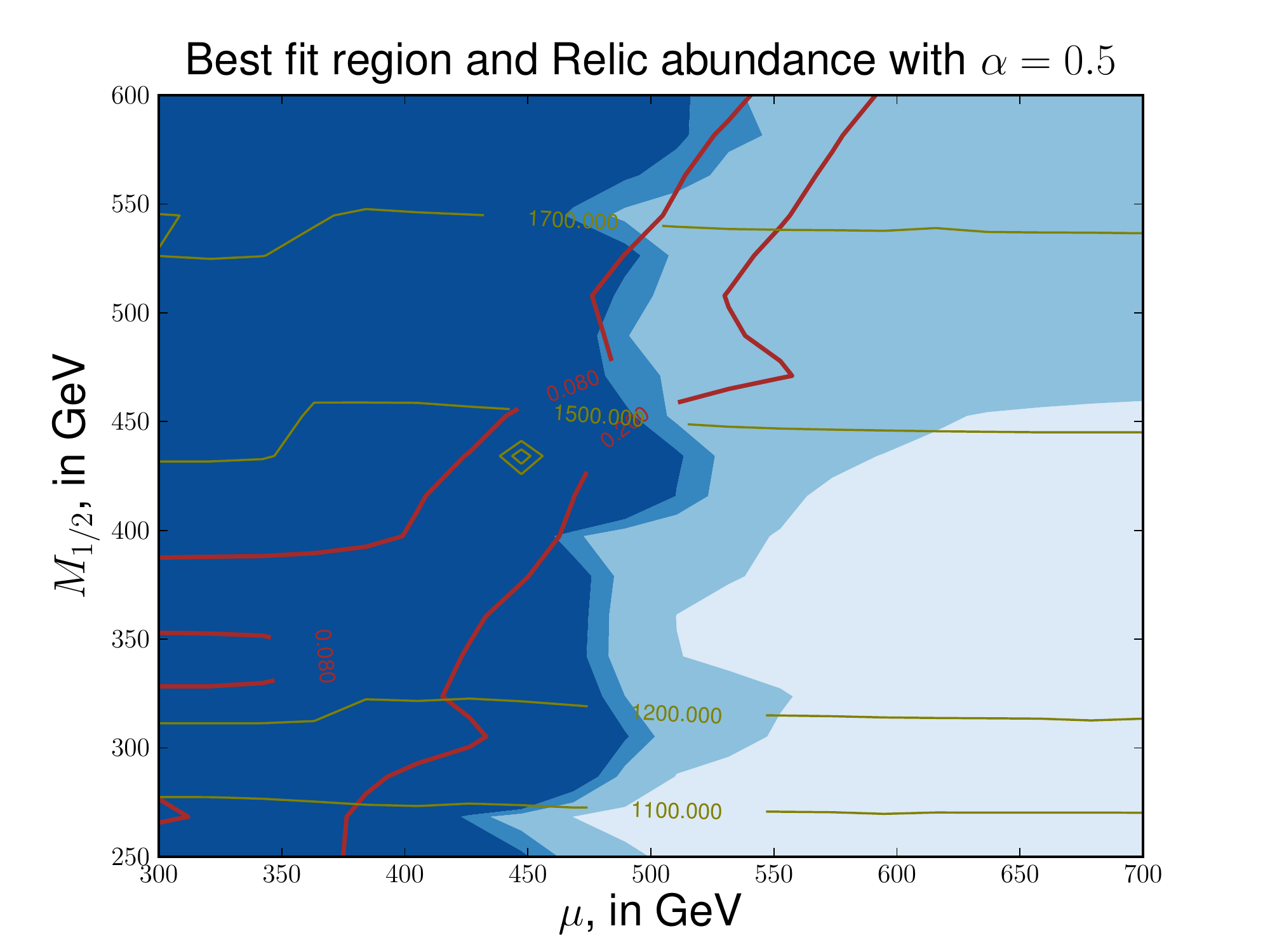}
\includegraphics[width=8cm]{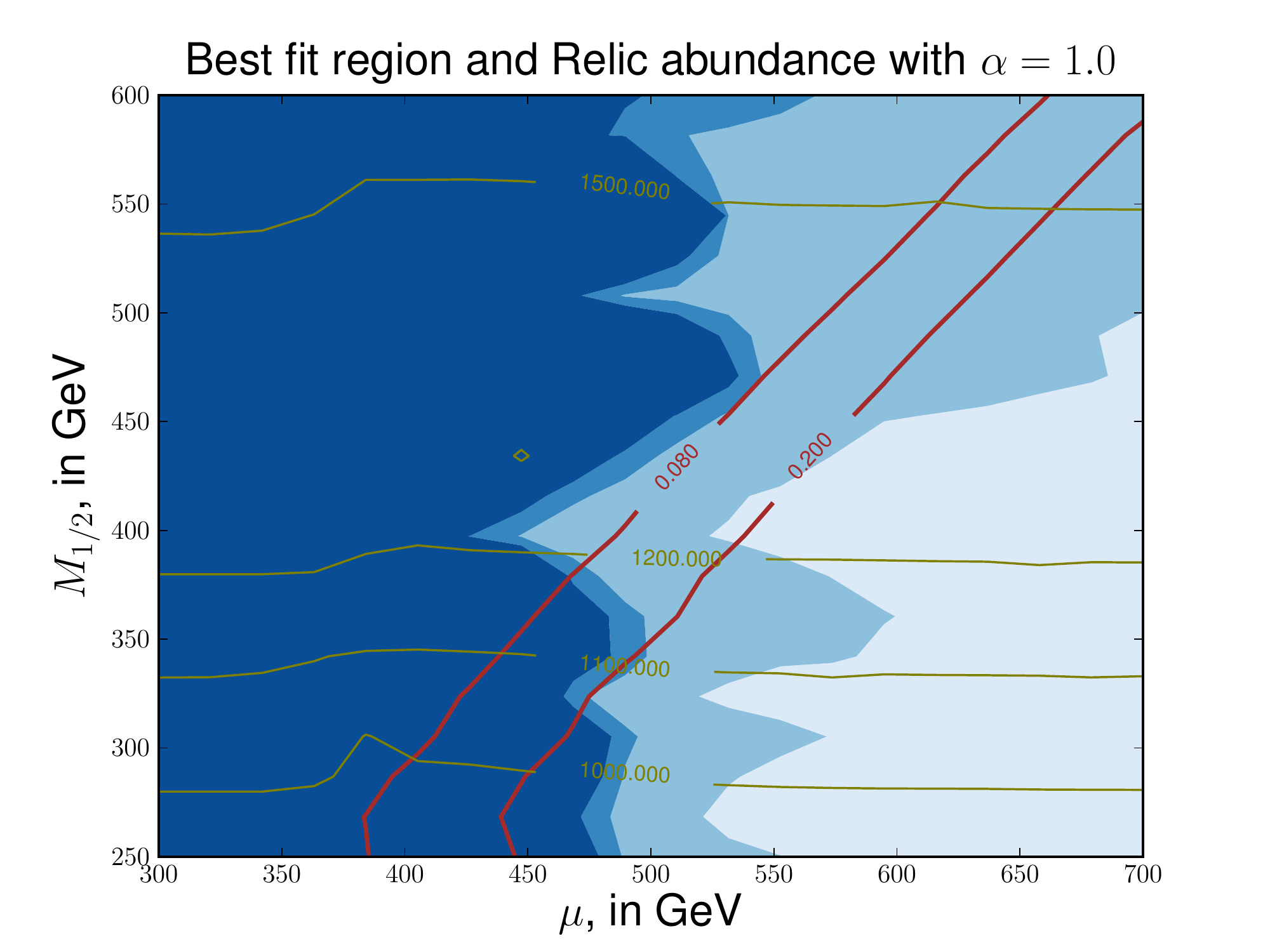}
\includegraphics[width=8cm]{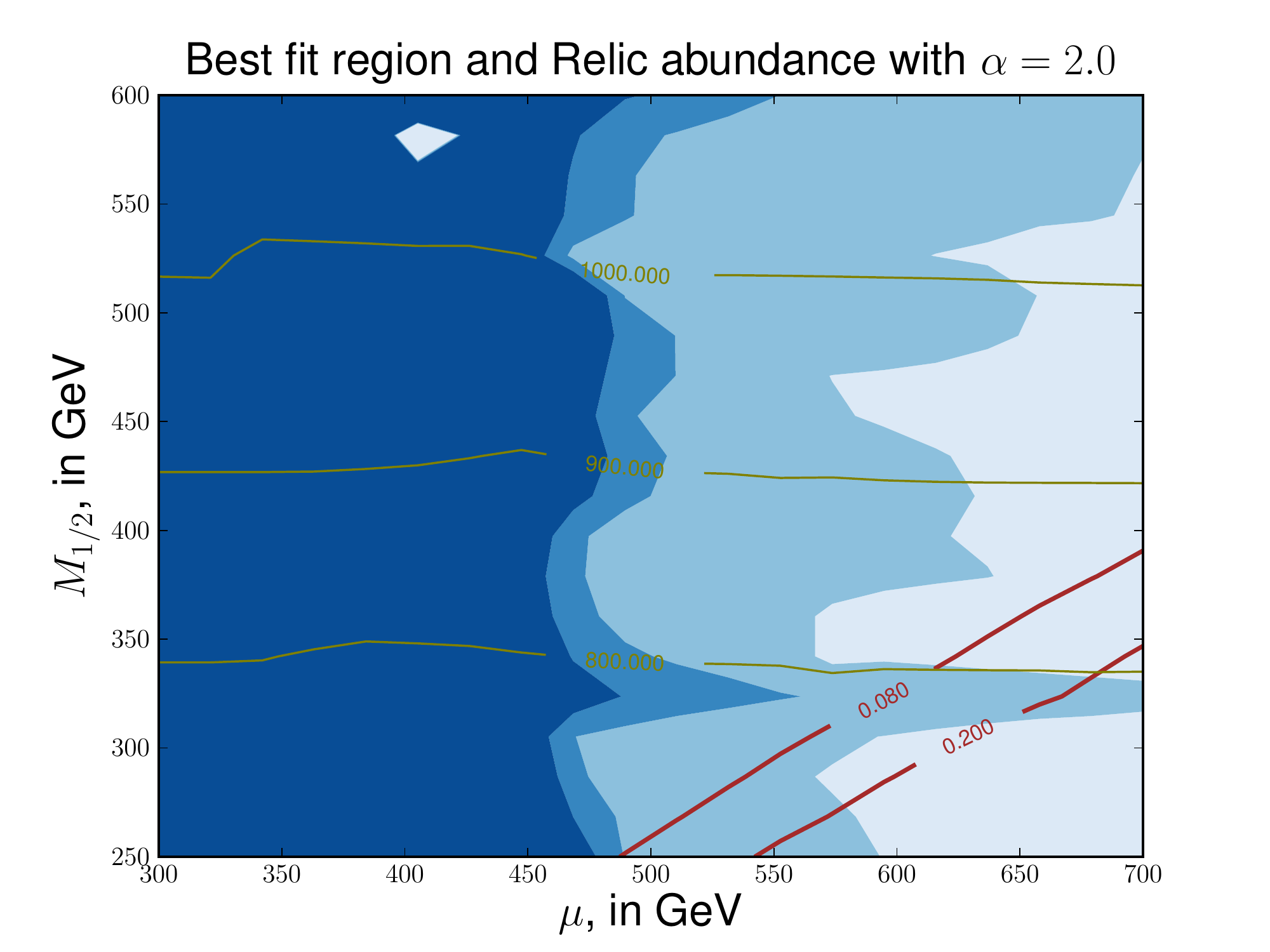}
\caption{\label{otheralpha} {\footnotesize {\bf Dependence on $\alpha$}: Best fit regions on a graph of $M_{1/2}$ versus $\mu$ in the case of $\alpha = 0.5, 1,$ and $2$. The region between the red lines gives $\Omega h^2 = 0.08 \, - \, 0.2$. The olive contours give the gluino mass. The blue contours (lightest to darkest) represent $\chi^2/dof = 1, 2.3, 3$ corresponding to 95\%, 90\%, and 68\% CLs, respectively. The figures show progressively better results as $\alpha$ is increased; the value of $\alpha = 1.5$ is shown in the main text as the benchmark point. The best fit is obtained for the case of $\alpha = 2$; however, the gluino mass is excluded in that case.}}
\end{center} 
\end{figure}

\end{appendices}

\clearpage
\newpage

\bibliography{bibliography}

\providecommand{\href}[2]{#2}\begingroup\raggedright\begin{thebibliography}{10}

\bibitem{Dimopoulos:1981yj}
S.~Dimopoulos, S.~Raby, and F.~Wilczek, ``{Supersymmetry and the Scale of
  Unification},'' {\em Phys.Rev.} {\bf D24} (1981)
1681--1683.

\bibitem{Dimopoulos:1981zb}
S.~Dimopoulos and H.~Georgi, ``{Softly Broken Supersymmetry and SU(5)},'' {\em
  Nucl.Phys.} {\bf B193} (1981)
150.

\bibitem{Ibanez:1981yh}
L.~E. Ibanez and G.~G. Ross, ``{Low-Energy Predictions in Supersymmetric Grand
  Unified Theories},'' {\em Phys.Lett.} {\bf B105} (1981)
439.

\bibitem{Sakai:1981gr}
N.~Sakai, ``{Naturalness in Supersymmetric Guts},'' {\em Z.Phys.} {\bf C11}
  (1981)
153.

\bibitem{Einhorn:1981sx}
M.~Einhorn and D.~Jones, ``{The Weak Mixing Angle and Unification Mass in
  Supersymmetric SU(5)},'' {\em Nucl.Phys.} {\bf B196} (1982)
475.

\bibitem{Marciano:1981un}
W.~J. Marciano and G.~Senjanovic, ``{Predictions of Supersymmetric Grand
  Unified Theories},'' {\em Phys.Rev.} {\bf D25} (1982)
3092.

\bibitem{Blazek:2001sb}
T.~Blazek, R.~Dermisek, and S.~Raby, ``{Predictions for Higgs and supersymmetry
  spectra from SO(10) Yukawa unification with $\mu > 0$},'' {\em
  Phys.Rev.Lett.} {\bf 88} (2002) 111804,
\href{http://www.arXiv.org/abs/hep-ph/0107097}{{\tt hep-ph/0107097}}.

\bibitem{Baer:2001yy}
H.~Baer and J.~Ferrandis, ``{Supersymmetric SO(10) GUT models with Yukawa
  unification and a positive $\mu$ term},'' {\em Phys.Rev.Lett.} {\bf 87}
  (2001) 211803,
\href{http://www.arXiv.org/abs/hep-ph/0106352}{{\tt hep-ph/0106352}}.

\bibitem{Blazek:2002ta}
T.~Blazek, R.~Dermisek, and S.~Raby, ``{Yukawa unification in SO(10)},'' {\em
  Phys.Rev.} {\bf D65} (2002) 115004,
\href{http://www.arXiv.org/abs/hep-ph/0201081}{{\tt hep-ph/0201081}}.

\bibitem{Tobe:2003bc}
K.~Tobe and J.~D. Wells, ``{Revisiting top bottom tau Yukawa unification in
  supersymmetric grand unified theories},'' {\em Nucl.Phys.} {\bf B663} (2003)
  123--140,
\href{http://www.arXiv.org/abs/hep-ph/0301015}{{\tt hep-ph/0301015}}.

\bibitem{Auto:2003ys}
D.~Auto, H.~Baer, C.~Balazs, A.~Belyaev, J.~Ferrandis, {\em et al.}, ``{Yukawa
  coupling unification in supersymmetric models},'' {\em JHEP} {\bf 0306}
  (2003) 023,
\href{http://www.arXiv.org/abs/hep-ph/0302155}{{\tt hep-ph/0302155}}.

\bibitem{Gogoladze:2011ce}
I.~Gogoladze, Q.~Shafi, and C.~S. Un, ``{SO(10) Yukawa Unification with mu $<$
  0},'' {\em Phys.Lett.} {\bf B704} (2011) 201--205,
\href{http://www.arXiv.org/abs/1107.1228}{{\tt 1107.1228}}.

\bibitem{Gogoladze:2011aa}
I.~Gogoladze, Q.~Shafi, and C.~S. Un, ``{Higgs Boson Mass from t-b-$\tau$
  Yukawa Unification},'' {\em JHEP} {\bf 1208} (2012) 028,
\href{http://www.arXiv.org/abs/1112.2206}{{\tt 1112.2206}}.

\bibitem{Badziak:2011wm}
M.~Badziak, M.~Olechowski, and S.~Pokorski, ``{Yukawa unification in SO(10)
  with light sparticle spectrum},'' {\em JHEP} {\bf 1108} (2011) 147,
\href{http://www.arXiv.org/abs/1107.2764}{{\tt 1107.2764}}.

\bibitem{Anandakrishnan:2012tj}
A.~Anandakrishnan, S.~Raby, and A.~Wingerter, ``{Yukawa Unification Predictions
  for the LHC},''
\href{http://www.arXiv.org/abs/1212.0542}{{\tt 1212.0542}}.

\bibitem{Anandakrishnan:2013cwa}
A.~Anandakrishnan and S.~Raby, ``{Yukawa Unification Predictions with effective
  "Mirage" Mediation},''
\href{http://www.arXiv.org/abs/1303.5125}{{\tt 1303.5125}}.

\bibitem{Ajaib:2013zha}
M.~Adeel~Ajaib, I.~Gogoladze, Q.~Shafi, and C.~S. Un, ``{A Predictive Yukawa
  Unified SO(10) Model: Higgs and Sparticle Masses},''
\href{http://www.arXiv.org/abs/1303.6964}{{\tt 1303.6964}}.

\bibitem{Moroi:1999zb}
T.~Moroi and L.~Randall, ``{Wino cold dark matter from anomaly mediated SUSY
  breaking},'' {\em Nucl.Phys.} {\bf B570} (2000) 455--472,
\href{http://www.arXiv.org/abs/hep-ph/9906527}{{\tt hep-ph/9906527}}.

\bibitem{Allahverdi:2012gk}
R.~Allahverdi, B.~Dutta, and K.~Sinha, ``{Successful Supersymmetric Dark Matter
  with Thermal Over/Under-Abundance from Late Decay of a Visible Sector
  Scalar},'' {\em Phys.Rev.} {\bf D87} (2013) 075024,
\href{http://www.arXiv.org/abs/1212.6948}{{\tt 1212.6948}}.

\bibitem{Allahverdi:2013noa}
R.~Allahverdi, M.~Cicoli, B.~Dutta, and K.~Sinha, ``{Non-thermal Dark Matter in
  String Compactifications},''
\href{http://www.arXiv.org/abs/1307.5086}{{\tt 1307.5086}}.

\bibitem{Allahverdi:2012wb}
R.~Allahverdi, B.~Dutta, and K.~Sinha, ``{Non-thermal Higgsino Dark Matter:
  Cosmological Motivations and Implications for a 125 GeV Higgs},'' {\em
  Phys.Rev.} {\bf D86} (2012) 095016,
\href{http://www.arXiv.org/abs/1208.0115}{{\tt 1208.0115}}.

\bibitem{Baer:2008eq}
H.~Baer and H.~Summy, ``{SO(10) SUSY GUTs, the gravitino problem, non-thermal
  leptogenesis and axino dark matter},'' {\em Phys.Lett.} {\bf B666} (2008)
  5--9,
\href{http://www.arXiv.org/abs/0803.0510}{{\tt 0803.0510}}.

\bibitem{Baer:2008yd}
H.~Baer, M.~Haider, S.~Kraml, S.~Sekmen, and H.~Summy, ``{Cosmological
  consequences of Yukawa-unified SUSY with mixed axion/axino cold and warm dark
  matter},'' {\em JCAP} {\bf 0902} (2009) 002,
\href{http://www.arXiv.org/abs/0812.2693}{{\tt 0812.2693}}.

\bibitem{Dermisek:2003vn}
R.~Dermisek, S.~Raby, L.~Roszkowski, and R.~Ruiz De~Austri, ``{Dark matter and
  $B_s \to \mu^{+} \mu^{-}$ with minimal SO(10) soft SUSY breaking},'' {\em
  JHEP} {\bf 0304} (2003) 037,
\href{http://www.arXiv.org/abs/hep-ph/0304101}{{\tt hep-ph/0304101}}.

\bibitem{Dermisek:2005sw}
R.~Dermisek, S.~Raby, L.~Roszkowski, and R.~Ruiz~de Austri, ``{Dark matter and
  $B_s \rightarrow \mu^+ \mu^-$ with minimal SO(10) soft SUSY breaking II},''
  {\em JHEP} {\bf 0509} (2005) 029,
\href{http://www.arXiv.org/abs/hep-ph/0507233}{{\tt hep-ph/0507233}}.

\bibitem{Baer:2008jn}
H.~Baer, S.~Kraml, S.~Sekmen, and H.~Summy, ``{Dark matter allowed scenarios
  for Yukawa-unified SO(10) SUSY GUTs},'' {\em JHEP} {\bf 0803} (2008) 056,
\href{http://www.arXiv.org/abs/0801.1831}{{\tt 0801.1831}}.

\bibitem{ArkaniHamed:2006mb}
N.~Arkani-Hamed, A.~Delgado, and G.~Giudice, ``{The Well-tempered
  neutralino},'' {\em Nucl.Phys.} {\bf B741} (2006) 108--130,
\href{http://www.arXiv.org/abs/hep-ph/0601041}{{\tt hep-ph/0601041}}.

\bibitem{ATLAS-CONF-2013-061}
``Search for strong production of supersymmetric particles in final states with
  missing transverse momentum and at least three b-jets using 20.1 fb−1 of pp
  collisions at sqrt(s) = 8 tev with the atlas detector.,'' Tech. Rep.
  ATLAS-CONF-2013-061, CERN, Geneva, Jun, 2013.

\bibitem{Chatrchyan:2013wxa}
{\bf CMS Collaboration} Collaboration, S.~Chatrchyan {\em et al.}, ``{Search
  for gluino mediated bottom- and top-squark production in multijet final
  states in pp collisions at 8 TeV},''
\href{http://www.arXiv.org/abs/1305.2390}{{\tt 1305.2390}}.

\bibitem{:2012gu}
{\bf CMS Collaboration} Collaboration, S.~Chatrchyan {\em et al.},
  ``{Observation of a new boson at a mass of 125 GeV with the CMS experiment at
  the LHC},'' {\em Phys.Lett.} {\bf B716} (2012) 30--61,
\href{http://www.arXiv.org/abs/1207.7235}{{\tt 1207.7235}}.

\bibitem{:2012gk}
{\bf ATLAS Collaboration} Collaboration, G.~Aad {\em et al.}, ``{Observation of
  a new particle in the search for the Standard Model Higgs boson with the
  ATLAS detector at the LHC},'' {\em Phys.Lett.} {\bf B716} (2012) 1--29,
\href{http://www.arXiv.org/abs/1207.7214}{{\tt 1207.7214}}.

\bibitem{Choi:2007ka}
K.~Choi and H.~P. Nilles, ``{The Gaugino code},'' {\em JHEP} {\bf 0704} (2007)
  006,
\href{http://www.arXiv.org/abs/hep-ph/0702146}{{\tt hep-ph/0702146}}.

\bibitem{Lowen:2008fm}
V.~Lowen and H.~P. Nilles, ``{Mirage Pattern from the Heterotic String},'' {\em
  Phys.Rev.} {\bf D77} (2008) 106007,
\href{http://www.arXiv.org/abs/0802.1137}{{\tt 0802.1137}}.

\bibitem{Baer:2006tb}
H.~Baer, E.-K. Park, X.~Tata, and T.~T. Wang, ``{Measuring Modular Weights in
  Mirage Unification Models at the LHC and ILC},'' {\em Phys.Lett.} {\bf B641}
  (2006) 447--451,
\href{http://www.arXiv.org/abs/hep-ph/0607085}{{\tt hep-ph/0607085}}.

\bibitem{Blazek:1995nv}
T.~Blazek, S.~Raby, and S.~Pokorski, ``{Finite supersymmetric threshold
  corrections to CKM matrix elements in the large $\tan \beta$ regime},'' {\em
  Phys.Rev.} {\bf D52} (1995) 4151--4158,
\href{http://www.arXiv.org/abs/hep-ph/9504364}{{\tt hep-ph/9504364}}.

\bibitem{Beringer:1900zz}
{\bf Particle Data Group} Collaboration, J.~Beringer {\em et al.}, ``{Review of
  Particle Physics (RPP)},'' {\em Phys.Rev.} {\bf D86} (2012)
010001.

\bibitem{hfag:2012-10-24}
{Heavy Flavor Averaging Group}. \url{http://www.slac.stanford.edu/xorg/hfag/},
  2012.

\bibitem{:2012ct}
{\bf LHCb Collaboration} Collaboration, R.~Aaij {\em et al.}, ``{First evidence
  for the decay Bs $\rightarrow$ mu+ mu-},''
\href{http://www.arXiv.org/abs/1211.2674}{{\tt 1211.2674}}.

\bibitem{Ade:2013ktc}
{\bf Planck Collaboration} Collaboration, P.~Ade {\em et al.}, ``{Planck 2013
  results. I. Overview of products and scientific results},''
\href{http://www.arXiv.org/abs/1303.5062}{{\tt 1303.5062}}.

\bibitem{Bernal:2007uv}
N.~Bernal, A.~Djouadi, and P.~Slavich, ``{The MSSM with heavy scalars},'' {\em
  JHEP} {\bf 0707} (2007) 016,
\href{http://www.arXiv.org/abs/0705.1496}{{\tt 0705.1496}}.

\bibitem{Crivellin:2012jv}
A.~Crivellin, J.~Rosiek, P.~Chankowski, A.~Dedes, S.~Jaeger, {\em et al.},
  ``{SUSY\_FLAVOR v2: A Computational tool for FCNC and CP-violating processes
  in the MSSM},''
\href{http://www.arXiv.org/abs/1203.5023}{{\tt 1203.5023}}.

\bibitem{Belanger:2010pz}
G.~Belanger, F.~Boudjema, A.~Pukhov, and A.~Semenov, ``{micrOMEGAs: A Tool for
  dark matter studies},'' {\em Nuovo Cim.} {\bf C033N2} (2010) 111--116,
\href{http://www.arXiv.org/abs/1005.4133}{{\tt 1005.4133}}.

\bibitem{James:1975dr}
F.~James and M.~Roos, ``{Minuit: A System for Function Minimization and
  Analysis of the Parameter Errors and Correlations},'' {\em
  Comput.Phys.Commun.} {\bf 10} (1975)
343--367.

\bibitem{Roszkowski:2009sm}
L.~Roszkowski, R.~Ruiz~de Austri, R.~Trotta, Y.-L.~S. Tsai, and T.~A. Varley,
  ``{Global fits of the Non-Universal Higgs Model},'' {\em Phys.Rev.} {\bf D83}
  (2011) 015014,
\href{http://www.arXiv.org/abs/0903.1279}{{\tt 0903.1279}}.

\bibitem{Kowalska:2013hha}
K.~Kowalska, L.~Roszkowski, and E.~M. Sessolo, ``{Two ultimate tests of
  constrained supersymmetry},'' {\em JHEP} {\bf 1306} (2013) 078,
\href{http://www.arXiv.org/abs/1302.5956}{{\tt 1302.5956}}.

\bibitem{Guchait:2011hj}
M.~Guchait, D.~Roy, and D.~Sengupta, ``{Probing a Mixed Neutralino Dark Matter
  Model at the 7 TeV LHC},'' {\em Phys.Rev.} {\bf D85} (2012) 035024,
\href{http://www.arXiv.org/abs/1109.6529}{{\tt 1109.6529}}.

\bibitem{Anandakrishnan:2013nca}
A.~Anandakrishnan, B.~C. Bryant, S.~Raby, and A.~Wingerter, ``{LHC
  Phenomenology of SO(10) Models with Yukawa Unification},''
\href{http://www.arXiv.org/abs/1307.7723}{{\tt 1307.7723}}.

\bibitem{Aprile:2012nq}
{\bf XENON100 Collaboration} Collaboration, E.~Aprile {\em et al.}, ``{Dark
  Matter Results from 225 Live Days of XENON100 Data},'' {\em Phys.Rev.Lett.}
  {\bf 109} (2012) 181301,
\href{http://www.arXiv.org/abs/1207.5988}{{\tt 1207.5988}}.

\bibitem{Cheung:2012qy}
C.~Cheung, L.~J. Hall, D.~Pinner, and J.~T. Ruderman, ``{Prospects and Blind
  Spots for Neutralino Dark Matter},''
\href{http://www.arXiv.org/abs/1211.4873}{{\tt 1211.4873}}.

\bibitem{Dutta:2013sta}
B.~Dutta, T.~Kamon, N.~Kolev, K.~Sinha, K.~Wang, {\em et al.}, ``{Top Squark
  Searches Using Dilepton Invariant Mass Distributions and Bino-Higgsino Dark
  Matter at the LHC},'' {\em Phys.Rev.} {\bf D87} (2013) 095007,
\href{http://www.arXiv.org/abs/1302.3231}{{\tt 1302.3231}}.

\bibitem{Han:2013gba}
T.~Han, Z.~Liu, and A.~Natarajan, ``{Dark Matter and Higgs Bosons in the
  MSSM},''
\href{http://www.arXiv.org/abs/1303.3040}{{\tt 1303.3040}}.

\bibitem{Aprile:2012zx}
{\bf XENON1T collaboration} Collaboration, E.~Aprile, ``{The XENON1T Dark
  Matter Search Experiment},''
\href{http://www.arXiv.org/abs/1206.6288}{{\tt 1206.6288}}.

\bibitem{Kamionkowski:1991nj}
M.~Kamionkowski, ``{Energetic neutrinos from heavy neutralino annihilation in
  the sun},'' {\em Phys.Rev.} {\bf D44} (1991)
3021--3042.

\bibitem{Griest:1989zh}
K.~Griest, M.~Kamionkowski, and M.~S. Turner, ``{Supersymmetric Dark Matter
  Above the W Mass},'' {\em Phys.Rev.} {\bf D41} (1990)
3565--3582.

\end{thebibliography}\endgroup

\bibliographystyle{utphys}


\end{document}